# The Theory of Thermodynamic Relativity

George Livadiotis* and David J. McComas

Department of Astrophysical Sciences, Princeton University, Princeton, NJ 08544, USA

**Abstract:** We introduce the theory of thermodynamic relativity, a unified theoretical framework for describing both entropies and velocities, and their respective physical disciplines of thermodynamics and kinematics, which share a surprisingly identical description with relativity. This is the first study to generalize relativity in a thermodynamic context, leading naturally to anisotropic and nonlinear adaptations of relativity; thermodynamic relativity constitutes a new path of generalization, as compared to the "traditional" passage from special to general theory based on curved spacetime. We show that entropy and velocity are characterized by three identical postulates, which provide the basis of a broader framework of relativity: (1) no privileged reference frame with zero value; (2) existence of an invariant and fixed value for all reference frames; and (3) existence of stationarity. The postulates lead to a unique way of addition for entropies and for velocities, called kappa-addition. We develop a systematic method of constructing a generalized framework of the theory of relativity, based on the kappa-addition formulation, that is fully consistent with both thermodynamics and kinematics. We call this novel and unified theoretical framework for simultaneously describing entropy and velocity "thermodynamic relativity". From the generality of the kappa-addition formulation, we focus on the cases corresponding to linear adaptations of special relativity. Then, we show how the developed thermodynamic relativity leads to the addition of entropies in nonextensive thermodynamics and the addition of velocities in Einstein's isotropic special relativity, as in two extreme cases, while intermediate cases correspond to a possible anisotropic adaptation of relativity. Using thermodynamic relativity for velocities, we start from the kappa-addition of velocities and construct the basic formulations of the linear anisotropic special relativity; e.g., the asymmetric Lorentz transformation, the nondiagonal metric, and the energy-momentum-velocity relationships. Then, we discuss the physical consequences of the possible anisotropy in known relativistic effects, such as, (i) matter-antimatter asymmetry, (ii) time dilation, and (iii) Doppler effect, and show how these might be used to detect and quantify a potential anisotropy.

## 1. Introduction

The development of thermodynamics was driven by our worldly experience with gasses, which are coupled through short-range, collisional interactions, and generally, reside in thermal equilibrium distributions. In contrast, space plasmas, from the solar wind and planetary magnetospheres to the outer heliosphere and beyond to interstellar and galactic plasmas, are quite different as particles have correlations and interact through longer-range electromagnetic interactions. Thus, space plasmas provide a natural laboratory for directly observing plasma particle distributions and for the observational ground truth in the development of a new and broader paradigm of thermodynamics. The journey to improve our understanding of the physical underpinnings of space thermodynamics has led to discovering new fundamental physics, including the concept of entropy defect [1], the generalization of the zeroth law of thermodynamics [2], and in this study, the development of thermodynamic relativity under a unified theoretical framework for describing both entropies and velocities. This is a novel theory, which should not be confused with some relativistic adaptation of thermodynamics, such as, the description of particle distributions and their thermodynamics in relativistic regimes of kinetic energies. Instead, it is a unification of two fundamental physical disciplines, those of thermodynamics and kinematics, which share a relativity description that is surprisingly identical. Velocity and entropy are basic physical variables of kinematics and thermodynamics, respectively. In an abstract description, velocity measures the change in position of a body through a motion, while entropy measures the corresponding change of information or order/disorder of this body. Thus, they appear to describe entirely different physical contexts. However, we show here that they share the same postulates and mathematical framework that was thought to be characteristic only of special relativity. Another common property is stationarity for both the velocity and entropy, which we upgraded here to be a new, formal, postulate. As shown in this study, the common postulates provide the basis that leads to a common mathematic formalism and unified relativity for velocities and entropies.

It has been nearly 60 years since the first observations of magnetospheric electrons, whose velocities unexpectedly deviated from the classical kinetic description of a Maxwellian distribution [3-5]. Since then, numerous observations of space plasma throughout the heliosphere have followed and repeatedly verified this specific non-Maxwellian behaviour [6-11]. Various empirical models of these distributions have been suggested, some simpler, some more mathematically complex, but all originating from the perspectives of generalized expressions, rather than physical first principles [6,12,13].

Empirical kappa distributions have been found to well describe the velocities of a plethora of space plasma particles, generalizing the classical Maxwellian distribution [14], by means of a parameter kappa, $\kappa$, that provides a measure of the shape of these distributions, in addition to the standard parameterization of temperature. The classical, Maxwellian distribution is included as the special kappa value of $\kappa \rightarrow \infty$.

Various names were given to the particle populations described by these distributions, with the most frequent being *suprathermal* and *nonthermal*. The term suprathermal was given to describe the non-Maxwellian distribution tails; as it was thought, the core of the distribution was still Maxwellian, with a suprathermal tail that was enhanced above the Maxwellian (e.g., [15-19]). Later, it was realized that both the Maxwellian core and the non-Maxwellian suprathermal tail are actually part of the same distribution, a kappa distribution (e.g., [9,10]. The nonthermal characterization came from the fact that the Maxwellian distribution describes particles in thermal equilibrium, thus non-Maxwellian distributions were thought to imply nonthermal particle populations. Now we understand that this was not accurate characterization; because how can a statistical distribution of particle velocities be nonthermal and simultaneously be parameterized by temperature? Temperature is the key-parameter of the zeroth law of thermodynamics that equalizes the inwards/outwards flow of heat when the particle system resides in thermal equilibrium. It was clear that a drastic change of classical statistical mechanics and thermodynamics was needed.

The first theoretical efforts came from the connection of kappa distributions with nonextensive statistical mechanics (e.g., [20-22] and [6]). This statistical framework is constructed on the basis of a generalized entropic formulation, called $q$-entropy [23,24], which includes the classical Boltzmann [25] – Gibbs [26] (BG) formulation as a special case, $q \to 1$. It is also called kappa entropy, as it is the entropy associated with kappa distributions; indeed, the maximization of the $q$-entropy, under the constraints of the canonical ensemble leads to the kappa distribution, where the kappa and $q$ parameters are trivially equated through $q = 1 + 1/\kappa$ [6,27]. In addition, Livadiotis and McComas [6,28] showed that this path connects the theory and formalism of kappa distributions with nonextensive statistical mechanics under consistent and equivalent kinetic and thermodynamic definitions of temperature.

In general, the maximized entropy leads to the canonical distributions (i.e., the Gibb's path [26]), but this does not mean that it counts as an origin of these distributions. In fact, this is simply a self-consistent derivation, as one can always find a specific entropy formulation that can lead to a certain distribution function when maximized [29-31]. The kappa distribution emerges within the framework of statistical mechanics by maximizing Tsallis entropy under the constraints of canonical ensemble. Nevertheless, this entropy maximization cannot be considered to be the origin of kappa distributions; both the entropic and distribution functions can be equivalently derived from each other. Therefore, the question still remains:

What is the thermodynamic origin of both the kappa distributions and their entropy? Or, equivalently: Are the kappa distributions and their entropy consistent with thermodynamics?

A misconception of the thermodynamic origin concerns the existence of mechanisms that can generate kappa distributions in plasmas. Some examples are: superstatistics [32–37], shock waves [38,39], turbulence [40-42], colloidal particles [43], interaction with pickup ions [44,45], pump acceleration mechanism [46], polytropic behavior [47,48]; Debye shielding and magnetic coupling [49-51]; (see also [10], Ch. 5, 6, 8, 10, 15, 16). While there are a variety of such mechanisms that can occur in space plasmas, thermodynamics ultimately determines if a particular distribution is allowed. Therefore, none of these mechanisms can explain whether kappa distributions (and their associated entropy) are consistent with thermodynamics.

In general, the origin of a particle distribution and its associated entropy, which are capable of physically describing particle systems, must be based on first principles of thermodynamics [1,2]. In order to derive the most generalized formulation that consistently represents entropy, we focus on the possible ways that the entropy of a system partitions into the entropies of the system's constituents (e.g., individual or groups of particles) [1,2,52-61]. The entropy partitioning has been approached in two ways, restricted and unrestricted [1]. The restricted way describes classical thermodynamics. According to this, the entropy is restricted to be an additive quantity, and systems can reach the classical thermal equilibrium, that is, a special stationary state interwoven with the following equivalent properties: (i) entropy is restricted to be additive, (ii) the formulation of entropy is given by the BG statistical framework [25,26], and (iii) the velocity distribution that maximizes this entropy within the constraints of the canonical ensemble is expressed by the Maxwell-Boltzmann formulation [14]. In contrast, the unrestricted way describes generalized thermodynamics. According to this, the entropy is not restricted by any addition rule, and systems can reach generalized thermal equilibrium, that is, any stationary state interwoven with the following properties: (i) kappa-addition of entropies; although entropy can be initially assumed to be unrestricted, the consistence of math naturally leads to a certain rule of addition, which stands as the most generalized way of entropy partitioning, that is, the kappa-addition of entropies [58,60]; (ii) the formulation of entropy is given by a general framework of nonextensive statistical mechanics such as the Tsallis entropy [23,54], and (iii) the velocity distribution that maximizes this entropy within the constraints of the canonical ensemble is given by the formulation of kappa distributions [6,29,30,61]. (Further details on classical/restricted vs. generalized/non-restricted thermodynamics can be found in Section 2 of [1].)

Therefore, the possible ways of entropy partitioning have a fundamental role in thermodynamics. Let the entropy of a system, composed from two parts A and B, be given as a function of their entropies, $S_A$ and $S_B$, respectively. This property of composability is due to the fact that entropy is the only macroscopic thermodynamic quantity that is well-defined for both stationary and non-stationary states; in contrast to temperature and related thermal variables, which are well-defined only at stationary states. Therefore, the entropy of the composed system, denoted by $A \oplus_\kappa B$, is $S_{A \oplus_\kappa B} = f(S_A, S_B)$. The symbol $\oplus_\kappa$ refers to the $\kappa$-addition, the mathematical formulation of the generalized portioning of entropies, which recovers the classical case of standard addition, $S_{A \oplus_\kappa B}(\kappa \to \infty) = S_A + S_B$; (recall that classical thermodynamics, which forces the partitioning of entropies to be additive, leads to the BG entropy and Maxwellian distribution [1,52]). Instead, our search shifted to the most generalized way of entropy partitioning, that is, the most generalized expression of a system's entropy as a function of the entropies of the system's constituents. We developed this expression through the concept of entropy defect [1].

Briefly, the entropy defect leads to the most generalized way of entropy partitioning that is consistent with the laws of thermodynamics. Entropy is a physical quantity that quantifies the disorder of a system. When a particle system resides in classical thermodynamic equilibrium, the entropy has a very simple way of being shared among the particles: it sums their entropies. However, when a particle system resides in the generalized thermodynamic equilibrium, such as space plasmas, this summation rule does not hold, as there is an additional term that reduces the total entropy, so that the total becomes less than the sum of the individual entropies. Particles in space plasmas move self-consistently with electromagnetic fields (e.g., Debye shielding [49-51], frozen in magnetic field [49]), which interact in ways that bind these particles together and produce correlations among particles. The existence of particle correlations adds order to the whole system, and thus, decreases its total disorder, or entropy. This concept is analogous to the mass defect that arises when nuclear particle systems are assembled: the total mass is less than the sum of the assembled masses because of the mass-energy spent in the fields, which bind the particles together (Figure 1).

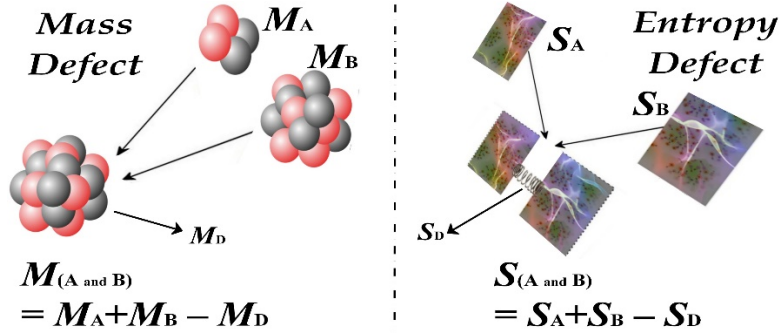

**Figure 1.** Schematic diagram of mass and entropy defects. Analogous to the mass defect ($M_D$) that quantifies the missing mass (energy) associated with assembling subatomic particles, the entropy defect ($S_D$) quantifies the missing entropy (order) associated with assembling space plasma particles. (Taken from [2])

The entropy defect describes how the entropy of the system partitions into the entropies of the system's constituents [1,2,45,52-54]. Take two, originally independent, constituents A and B with entropies $S_A$ and $S_B$, respectively, which are assembled into a composed system of entropy $S_{A \oplus B}$, with additional correlations developed between the two constituents. Then, the order induced by the developed correlations causes the system's combined entropy to decrease, and thus become less than the simple sum of entropies of the constituents, $S_{A \oplus B} - (S_A + S_B) < 0$. The missing entropy defines the entropy defect $S_D \equiv (S_A + S_B) - S_{A \oplus B}$. Note that the setup of the two subsystems A and B is a thermodynamically close system. As such, the entropy of the system before its composition, that is the summation of the entropies of two independent systems, $S_A + S_B$, and the entropy after the composition plus the entropy defect that measures the entropy spent on the additional correlations developed once the total system is composed, $S_{A \oplus B} + S_D$, are equal, i.e., $S_{A \oplus B} = S_A + S_B - S_D$. Moreover, the exact expression of the entropy defect was shown to be $S_D = \frac{1}{\kappa} \cdot S_A \cdot S_B$ [1,2], with $1/\kappa$ measuring the magnitude of the interconnectedness (i.e., correlations) among the system's constituents, which is causing the defect. The total entropy of the system is determined by a nonlinear expression of the constituent entropies, formulating a kappa-dependent addition rule for entropies partitioning, simply called, kappa-addition,

$$S_{A \oplus B} \equiv S_A \oplus_\kappa S_B = S_A + S_B - \tfrac{1}{\kappa} \cdot S_A \cdot S_B . \tag{1}$$

The entropy partitioning consistent with this addition rule leads to the specific formula of q- or kappa entropy [54-59]. The partitioning in Eq.(1) follows the simple entropy defect, while in its most generalized version, the entropy defect can be expressed through an arbitrary positive and increasingly monotonic function $H$ (see the full description of $H$ in the next section):

$$H(S_{A \oplus B}) \equiv H(S_A \oplus_\kappa S_B) = H(S_A) + H(S_B) - \tfrac{1}{\kappa} H(S_A) H(S_B) . \tag{2}$$

We showed [1,2], that the entropy defect can be determined from three basic axioms, which must hold for any partitioning function $H$: (1) Separability $S_\mathrm{D}(S_\mathrm{A},S_\mathrm{B}) \propto g(S_\mathrm{A}) \cdot h(S_\mathrm{B})$; (2) Symmetry, $S_\mathrm{D}(S_\mathrm{A},S_\mathrm{B}) = S_\mathrm{D}(S_\mathrm{B},S_\mathrm{A}) \propto g(S_\mathrm{A}) \cdot g(S_\mathrm{B})$; and (3) Upper boundedness, i.e., the existence of an upper limit of any entropy value, $S < S_{\max}$, where the upper limit is, in general a function of $\kappa$, i.e., $S_{\max} = S_{\max}(\kappa)$; e.g., the simple case of $H(S) = S$ gives $S_{\max} = \kappa$. Also, the composition of entropies under the addition of Eq.(2) can lead to a stationary state; namely, if $S_\mathrm{A}$ and $S_\mathrm{B}$ are the entropies of a stationary state, the total system is also residing in a stationary state with entropy $S_{\mathrm{A+B}}$ [58,60].

The addition rule of entropies, which is implied with the formulation of entropy defect, has some specific algebra [2]. This includes the transitive and symmetric properties, which are fundamental for the zero[th] law of thermodynamics. The zero[th] law of thermodynamics is naturally a transitive thermodynamic property of systems (a relation R on a set X is transitive if, for all elements A, B, C in X, whenever R relates A to B and B to C, then R also relates A to C). In addition, the zero[th] law of thermodynamics is also a symmetric property, i.e., If A is in a stationary state (generalized thermal equilibrium) with B, then B is in the same stationary state with A. The transitive and symmetric properties of the zero[th] law of thermodynamics conjure the matter of connections between systems. Indeed, the way that C is connected to A and B provides all the information on how A and B are connected. If, for instance, C is in a stationary state with A and with B, then, A and B are together in a stationary state.

The journey through understanding generalized thermodynamics led us to realize that the physical framework of entropic values also extends beyond the context of thermodynamics. Indeed, we note the following characteristics of any arbitrary entropy value $S$: (1) a non-standard addition rule and connection between observers; (2) existence of upper limit $S_{\max}$, so that for any entropy $S < S_{\max}$; and (3) existence of stationarity for any entropy $S$ (still less than $S_{\max}$). We cannot say strongly enough: these three characteristics are identical to the two traditional postulates of special relativity for velocities plus the trivial assumption of the existence of inertial frames.

In this study we use these parallel conditions to develop a unified relativity framework for both thermodynamics and kinematics, which we call, thermodynamic relativity. The purpose of this paper is to shed light on the relativity of both the entropies and velocities, examine their similarities, and finally, construct a unified framework of thermodynamic relativity. First, we apply thermodynamic relativity to entropies and show the consequences in statistical mechanics. Next, we apply thermodynamic relativity to velocities, show that this produces a naturally derived anisotropic special theory of relativity, and discuss some of the important consequences of this relativistic kinematics. For simplicity, here we consider linear motion and the 1-dimensional velocity, determining the effects of relativity in spacetime that involves one spatial dimension, the one of motion; for this, the 1D-velocity reduces to the characterization of speed.

The paper is organized as follows. Section 2 provides the general formulation of entropy partitioning, as determined by the entropy defect. Section 3 develops the postulates of the relativity of entropy: (1) principle of relativity for entropy; (2) existence of a fixed and invariant entropy, that is, an upper limit of entropy values; and (3) existence of stationary entropy. Section 4 revisits the relativity of velocities and restates and discusses its postulates: (1) all frames of reference are equivalent; (2) existence of a fixed and invariant speed, that is, an upper limit of speeds; and (3) existence of stationary velocity. Section 5 unifies the two relativities in one framework, thermodynamic relativity for entropies and velocities. This is shown to be a naturally derived anisotropic version of special relativity. We study the two extreme cases, that is, the isotropy (Einstein's special relativity) and maximum anisotropy (nonextensive thermodynamics), and show how these can be included in a single unified description. Section 6 focuses on the kinematics, i.e., the anisotropy in the speed of light, the velocity addition, the asymmetric matrix of the Lorentz transformation, the non-diagonal norm, the energy-momentum and energy-velocity equations. Section 7 examines the physical consequences of, and possibility for measuring, the anisotropy within the unified framework of thermodynamic relativity, focusing on the (i) matter-antimatter asymmetry, (ii) time dilation, and (iii) Doppler effect. Finally, Section 8 summarizes and discusses the conclusions. The supplementary material covers aspects of the formulation of the relativity framework for entropies and velocities.

## 2. Entropy defect – General formulation

The entropy partitioning formulates the addition rule that includes the entropy defect, as shown in Eq.(1). In general, this is expressed in terms of a function of the involved entropies, i.e., $H(S_A)$, $H(S_B)$, $H(S_{A\oplus B})$. Then, the entropy partitioning is formulated through a partitioning function $H$; rewriting Eq.(2),

$$S_{A\oplus B} \equiv S_A \oplus_\kappa S_B = H^{-1}\left[H(S_A) + H(S_B) - \tfrac{1}{\kappa}H(S_A)H(S_B)\right]. \tag{3}$$

This is the most general formulation of the entropy partitioning consistent with thermodynamics as shown by [58,60] for stationary systems, and then shown in general, even for non-stationary systems, through the path of the entropy defect by [1,2].

The partitioning function $H=H(S)$ is not uniquely determined. The properties characterizing this function are: (i) $H \geq 0$, where the zero holds at $S=0$; the non-negativity comes from the requirement of $H$ to equal $S$ in the classical limit (see property (v) below); (ii) $H(0)=0$, because adding zero entropy must have zero change in total entropy; indeed, setting $S_B = 0$ in Eq.(2), requiring that $H(S_{A\oplus B}) = H(S_A)$, we obtain $H(0) \cdot [1 - \tfrac{1}{\kappa}H(S_A)] = 0$ for any $S_A$, thus $H(0)=0$; (iii) $H'(0) = 1$, because $H'(0)$ appears always in a ratio with kappa, thus its value is arbitrary and can be absorbed into the kappa; indeed, for small entropies, we have $H(S) \cong H(0) + H'(0) \cdot S = H'(0) \cdot S + O(S^2)$, and the $H$-partitioning in Eq.(3) leads to $S_D \equiv S_A + S_B - S_{A\oplus B} = H'(0) \cdot \tfrac{1}{\kappa} S_A S_B + O(S_A^2 S_B) + O(S_A S_B^2)$, where the square term should be identical to $S_\text{D} = \frac{1}{\kappa} S_\text{A} S_\text{B}$, thus $H'(0) = 1$; (iv) $H$ is monotonically increasing, because of (iii) and that its inverse $H^{-1}(S)$ must be defined; (v) if $H$ is kappa dependent, then at the classical limit where $\kappa \to \infty$, it must reduce to the identity function $H(S)=S$ [50,58]; and (vi) the produced entropy defect $S_\text{D}$ must be positive, i.e., $S_\text{D} \equiv S_\text{A} + S_\text{B} - S_{\text{A} \oplus \text{B}} > 0$.

For any function $H$ following these properties, the entropy defect leads to the whole structure of thermodynamics [1], deriving the entropy [45,53], its statistical equation [54], its thermodynamic properties [2], the stationary state characterized by the canonical distribution function [1,50], and the connection of entropy and temperature, which are given by the thermodynamic definitions of temperature and kappa [53].

## 3. Postulates of the relativity of entropy

The *H*-partitioning of entropies and the corresponding kappa-addition follows three fundamental postulates that lead to the relativity of entropy: (1) no privileged reference frame with zero entropic value (principle of relativity for entropic values); (2) existence of an upper limit of entropic values; and (3) existence of stationarity of entropic values. The first two postulates follow the classical paradigm, while the third was considered a trivial condition in Einstein's special relativity for velocities, but is clearly necessary for entropies.

### *3.1. First Postulate: Principle of relativity for entropy*

We present the relative nature of entropy through the following three arguments: 1) Measurements of entropy come always via differences; 2) The definition of entropy as an absolute measure is simply by construction; 3) The definition entropy as a relative measure has already been expressed and studied (see below the Kullback–Leibler definition). Namely:

1) The entropy of a system is only ever obtained as an entropy difference, that is, the entropy of the system measured from a different reference state or frame; here, the thermodynamic reference frame is the stationary state of the observer, from which the entropy of some other body is measured. For instance, let the exoentropic chemical reaction, A→B, where an amount of entropy $S_{\text{B,A}}$ is released; then, the reverse reaction B→A is endoentropic, where an amount of entropy $S_{\text{A,B}}$ is absorbed. Now, if the entropies of A and B are measured from a different entropic level, let this be O (that is, the observer), then, we have that $S_{\text{B,A}} = S_{\text{B,O}} - S_{\text{A,O}}$.

2) The entropy is a relative physical quantity in that it can be measured only with respect to another reference frame. Thus, seeking an absolute value of entropy requires an arbitrary definition of entropic zero. Specifically, the third law of thermodynamics states that the entropy at zero temperature is a well-defined constant [62] that can be set to zero [63; 64]. However, like the false assumption of the existence of aether, absolute vacuum was taken as the absolute reference frame for measuring entropy. This frame is characterized by the least possible entropy, which is set as the absolute zero of entropy. The vacuum energy that characterizes the hypothetical aether is the zero-point energy – that is, the energy of the system at a temperature of absolute zero, and thus, at zero entropy. The classical perception of the aether is that it is immobile, i.e., the preferred reference frame of zero speed, but also, of zero entropy. Einstein, however, cleared up this misconception, as the absolute vacuum is characterized not just by immobility but also by nonexistence [65]. In particular, he did not interpret the zero metric as a realistic solution of his field equations, but rather as a mathematical possibility that has no physical significance.

3) In the perspective of relativity, the entropy of a system can only be determined from a reference frame in which the entropy is measured. The expression of entropy of a system through its probability distribution (e.g., the classical BG [26] or other generalized formulations such as the $q$-entropy [23,24], which is equivalent to kappa entropy [29,50,54]) constitutes an absolute definition. Instead, the relative entropy is expressed as a combination of both the probability distributions of the two difference systems. The most rigorous expression of entropy is through the information measure, called also "surprisal". Information quantifies the amount of surprise, through a specific function of the probability distribution, called information measure, while the entropy is defined as the expectation of the information measure, or, expected surprise [66,67]. Therefore, entropy measures the average amount of information needed to represent an event drawn from a probability distribution for a random variable. In its relative determination, the entropy of a system measured from a reference frame is expressed by the expected surprise of that system as measured from that reference frame (also called, Kullback–Leibler entropy difference) [68,69]. When measuring your own relative entropy, it would turn out to be zero: indeed, there is no expected surprise to be measured in your own system. (For the quantitative expression of relative entropy, see: [69], and Supplementary, Section A.)

For a system A with entropy $S_A$, it is implied that this is measured from some reference frame O (e.g., a laboratory), thus we note it as $S_{A,O}$. The entropy of another system B, measured from O is $S_{B,O}$. The connection of the two systems A and B leads to the measurement of entropy of B from A, $S_{B,A}$, and the measurement of its "inverse", the measurement of entropy of A from B, $S_{A,B}$. The mathematical properties of (i) commutativity between $S_{B,A}$ and $S_{A,B}$, and (ii) associativity between $S_{A,O}$, $S_{B,O}$, and $S_{A,B}$, reflect the physical properties of symmetry and transitivity, respectively, which characterize the zeroth law of thermodynamics [2], and the corresponding relationships are determined by the kappa addition shown in Eq.(3), i.e., (i) $S_{B,A} \oplus_\kappa S_{A,B} = 0$ with $S_{A,B} = \overline{S}_{B,A}$, and (ii) $S_{A,O} = S_{A,B} \oplus_\kappa S_{B,O}$ (Supplementary, Section A).

The connection between the entropies measured in different reference frames is described by the kappa addition. Given the function $H(S)$, we can conclude that the kappa addition forms a mathematical group on the set of entropies. This is shown through the following algebra, which generalizes the steps and properties developed in [2]: (i) Closure: For any two entropic values, $S_A$ and $S_B$, belonging to the set of possible entropies, $\Omega_S$, their addition belongs also to $\Omega_S$, $S_A, S_B \in \Omega_S \Rightarrow S_{A \oplus B} \in \Omega_S$. (ii) Identity: if $S_B = 0$, then for any $S_A \in \Omega_S$, $S_{A \oplus B} = S_A$. (iii) Inverse: for any $S_A$, there exists its inverse element with entropy $\bar{S}_A$, for which $S_A \oplus_\kappa \bar{S}_A = 0$, hence, we find that $H(\bar{S}_A) = -H(S_A)/[1 - \frac{1}{\kappa} H(S_A)]$. This defines the $\kappa$-subtraction of two elements B and A, that is, the $\kappa$-addition with the inverse of subtrahend, $S_B \oplus_\kappa \bar{S}_A$. The entropy of B measured by the reference frame of A, $S_{B,A}$, is determined by their subtraction, $S_{B,A} = S_B \oplus_\kappa \bar{S}_A$. (iv) Associativity: The entropy of B measured by A, $S_{B,A}$, can be expressed by the $\kappa$-addition of the entropy of B measured by C, $S_{B,C}$, and the entropy of C measured by A, $S_{C,A}$, i.e., $S_{B,A} = S_{B,C} \oplus_\kappa S_{C,A}$, or $H(S_{B,A}) = H(S_{B,C}) + H(S_{C,A}) - \frac{1}{\kappa} H(S_{B,C}) \cdot H(S_{C,A})$. (v) Commutativity: Additionally, the group is abelian, since the addition function is symmetric, $S_A \oplus_\kappa S_B = S_B \oplus_\kappa S_A$.

In summary, the first postulate states: *There is no absolute reference frame in which entropy is zero.* On the contrary, entropy is a relative quantity, connected with the properties of symmetry and transitivity, broadly defined under a general addition rule that forms a mathematical group on the set of allowable entropic values.

### 3.2. Second Postulate: Existence of a fixed and invariant entropy, upper limit of entropy values.

Adding two systems A and B together into a composed system $A \oplus B$ requires the total entropy of the composed system to be at least as large as any of component's entropies, $S_{A \oplus B} \geq S_A$ (see the strict proof consistent with thermodynamics in [1,2]); then, $H(S_{A \oplus B}) \geq H(S_A)$ ($H$: monotonically increasing function), leading to $1 - \frac{1}{\kappa} H(S_A) \geq 0$, defining an upper limit $c_S$ of entropy values $S$ (dropping the A and B subscripts):

$$S \leq H^{-1}(\kappa) \equiv c_S. \tag{4}$$

The upper limit constitutes an invariant and fixed value. It is invariant because remains the same, independent of the reference frame: If $c_S$ equals the entropy of B measured from C, $S_{B,C} = c_S$, then, applying $S_{B,A} = S_{B,C} \oplus_\kappa S_{C,A}$, we find $S_{B,A} = c_S$, i.e., $c_S$ also equals the entropy of B measured from A, a result that is independent of the entropy difference between the observing frames C and A, i.e., $S_{C,A}$ or $S_{A,C} = \bar{S}_{C,A}$. It is also fixed, because it determines a fixed point in the difference equation of entropy addition: The entropy of the system at the $i^{th}$ iteration, $S_i$, changes to $S_{i+1}$ when adding an entropy $\sigma$, according to $H(S_{i+1}) = H(S_i) + H(\sigma) - \frac{1}{\kappa} H(S_i) \cdot H(\sigma)$. Then, the entropic value $S_i = c_S$ is a fixed point (i.e., $S_{i+1} = S_i$).

The upper limit also recovers the entropic units of kappa. The nonlinear entropic relations typically have the entropies appear as unitless, that is, each entropic value is silently divided by the Boltzmann constant $k_B$. For example, we consider the case of the simple entropy defect, $S_{A \oplus B} = S_A + S_B - \frac{1}{\kappa} \cdot S_A \cdot S_B$, with extensive entropy relationship $S_\infty = -\kappa \cdot \ln(1 - \frac{1}{\kappa}S)$, which corresponds to the identity partitioning function $H(S)=S$. We observe that entropy appears always as a ratio of its value divided by kappa, $S/\kappa$. Indeed, if we set $\chi \equiv S/\kappa$, then there is no need for any assumption in regards to the units, e.g., $\chi_{A \oplus B} = \varphi(\chi_A, \chi_B) \equiv \chi_A + \chi_B - \chi_A \chi_B$, $\chi_\infty = -\ln(1-\chi)$. Consequently, the value of kappa can be set to have entropy units, i.e., $k_B$. Furthermore, more general $H$ functions can be written as $H(S) = S \cdot g(S/\kappa)$, so that the functional $1 - \frac{1}{\kappa}H(S)$ that appears in the $H$-partitioning involves only the ratio $S/\kappa$, i.e., $\frac{1}{\kappa}H(S_{A \oplus B}) = \varphi\left(\frac{1}{\kappa}H(S_A), \frac{1}{\kappa}H(S_B)\right)$. The upper limit is given by $\frac{1}{\kappa}H(c_S) = 1$ or $c_S \propto \kappa$, meaning that both $c_S$ and $\kappa$ have the same dimensions as $k_B$.

Therefore, the second postulate states: *There exists a nonzero, finite entropy value, which remains fixed (i.e., constant for all times) and invariant (i.e., constant for all observers), thus, it has the same value in all stationary frames of reference, and constitutes the upper limit of any entropy.*

### 3.3. Third Postulate: Existence of stationary entropies

Once the two systems A and B are connected, energy and entropy are allowed to flow and be exchanged, leading to a state of the composed system where the total entropy is expressed as a function of the individual original entropies $S_A$ and $S_B$ (a property called composability). Specifically, this function is formulated by the $H$-partitioning of Eq.(3), or the $\kappa$-addition of entropies. It has been shown that the $H$-partitioning constitutes the most general formalism that corresponds to stationarity [58,60]. Namely, for variations of the constituents' entropies, $S_A$ and $S_B$, in a way that the total entropy remains invariant, i.e., $S_{A \oplus B} = const.$ or $dS_{A \oplus B} = 0$, then, the total entropy is given by the kappa addition as in Eq.(3), $\frac{1}{\kappa}H(S_{A \oplus B}) = \varphi\left[\frac{1}{\kappa}H(S_A), \frac{1}{\kappa}H(S_B)\right]$.

Once the composed system resides in a stationary state, then a temperature can be thermodynamically defined. The thermodynamic definition of temperature comes from the relationship between entropy and internal energy $U$, $1/T \equiv \partial S_\infty / \partial U$, though, the involved entropic quantity $S_\infty$ behaves exactly as the entropy $S$ at the classical case of thermal equilibrium, $\kappa \to \infty$. The extensive measure of entropy, noted with $S_\infty$, is a mathematical quantity that becomes physically meaningful for systems residing in stationary states. This is because it serves as the connecting link of the actual entropy $S$ with the temperature $T$, which it can only be meaningful in stationary states (that is, generalized thermal equilibrium). The extensive measure $S_\infty$ has the units of $S$ and coincides with the classical BG entropy at the limit of $\kappa \to \infty$,

$$S_\infty = \ln[1 - \tfrac{1}{\kappa}H(S)]^{-\kappa}. \tag{5}$$

This relationship can be derived as follows: The $H$-partitioning can be written in the product form:

$$[1 - \tfrac{1}{\kappa}H(S_{A \oplus B})] = [1 - \tfrac{1}{\kappa}H(S_A)] \cdot [1 - \tfrac{1}{\kappa}H(S_B)]. \tag{6}$$

We observe that the logarithm of the quantities $1-\frac{1}{\kappa}H$ behave extensively, i.e., of this relationship $\ln[1-\frac{1}{\kappa}H(S_{A\oplus B})] = \ln[1-\frac{1}{\kappa}H(S_A)] + \ln[1-\frac{1}{\kappa}H(S_B)]$ is extensive. Then, the quantity $A \cdot \ln[1-\frac{1}{\kappa}H(S)]$ is extensive, while the constant A is taken as $A=-\kappa$ so it coincides with $S$ at the limit of $\kappa \to \infty$ (recall that at this limit $H$ is the identity function). Hence, we conclude with Eq.(5).

The relationship can be also shown through infinitesimal variations. In particular, we derive the change of the system's entropy $dS$, once an originally independent amount of entropy $dS_\infty$ is added to its initial entropy $S$, i.e., $S + dS = S \oplus_\kappa dS_\infty$. Setting $S_A \to S$, $S_B \to dS_\infty$, then $S_{A \oplus B} \to S + dS$, in Eq.(6), and considering $H(S+dS) = H(S) + H'(S)dS$, and $H'(0) = 1$, we find $dS_\infty = \{H'(S)/[1-\frac{1}{\kappa}H(S)]\} \cdot dS$, leading to the extensive measure of entropy $S_\infty$ given by Eq.(5).

The extensive measure $S_\infty$ depends on temperature, but not on kappa. In fact, it is given by the Sacker-Tetrode equation, $S_\infty = \frac{1}{2}d \cdot \ln T + const$. Then, the equation that determines thermodynamically the temperature is the same, independent of kappa, thus, as in the case of $\kappa \to \infty$, i.e., $1/T = \partial S_\infty / \partial U$, or

$$\frac{1}{T} \equiv \frac{\partial S_\infty}{\partial U} = \frac{\partial \ln[1-\frac{1}{\kappa}H(S)]^{-\kappa}}{\partial U} = \frac{H'(S)}{1-\frac{1}{\kappa}H(S)} \cdot \frac{\partial S}{\partial U}. \qquad (7)$$

Finally, stationarity is interwoven with the zeroth law of thermodynamics. If a system A is stationary for a reference frame O it will be stationary for any other system O′, which is also stationary with O. This comes from the zeroth law of thermodynamics, which is associated with the properties of transitivity and symmetry. The properties of symmetry and transitivity connect the entropies $S$ of systems, both for stationary and nonstationary states. Then, the zero[th] law of thermodynamics can be stated in terms of entropy difference: "*If a body C measures the entropies of two other bodies, A and B, $S_{A,C}$ and $S_{B,C}$, then, their combined entropy, $S_{A,C}$ and $S_{B,C}$, is measured as the connected A and B entropy, where the H-partitioning is involved in all the entropy measurements*" [2]. In particular, the law's transitivity states that $S_{A,O} = S_{A,O'} \oplus S_{O',O}$; the corresponding extensive measures replace the kappa addition with the standard sum, i.e., $S_{\infty A,O} = S_{\infty A,O'} + S_{\infty O',O}$. Then, the stationarity of O′ (from O) leads to $\partial S_{\infty A,O} = \partial S_{\infty A,O'}$, while the conservation of energy leads to a similar equation holds for internal energy, $\partial U_{A,O} = \partial U_{A,O'}$. Therefore, the thermodynamic definition of temperature of A is equivalent for all stationary observers (O or O′),

$$\frac{\partial S_{\infty A,O}}{\partial U_A} = \frac{\partial S_{\infty A,O'}}{\partial U_A} \equiv \frac{1}{T}. \qquad (8)$$

On the other hand, the law's symmetry between two connected systems A and B states that $S_{A,B} \oplus S_{B,A} = 0$, or equivalently, $S_{\infty A,B} + S_{\infty B,A} = 0$, thus $\partial S_{\infty A,B} = -\partial S_{\infty B,A}$. Also, the energy requires $\partial U_A = -\partial U_B$, hence,

$$\frac{\partial S_{\infty A,B}}{\partial U_A} = \frac{\partial S_{\infty B,A}}{\partial U_B} \equiv \frac{1}{T}. \qquad (9)$$

Finally, the third postulate states: *If a system is stationary for a reference frame O, it will be stationary for all reference frames that are stationary for O.*

## 4. Relativity of Velocity

Here we revisit and discuss the postulates of special relativity [70]: (1) First postulate (principle of relativity): The laws of physics take the same form in all inertial frames of reference; (2) Second postulate (invariance of *c*): the speed of light in free space has the same value in all inertial frames of reference; and (3) Third postulate, which has been added to express the necessity of the existence of stationarity. We show the surprising result that they are parallel to and essentially identical to those of thermodynamics.

### *4.1. First Postulate: All frames of reference are equivalent*

Inertial reference frames are systems with a constant bulk velocity. Observers define systems in different reference frames, which here are considered to be inertial. The term inertial here is identical to "stationary" but referring to the velocity space. Thus, throughout the paper we can characterize these frames as stationary, and examine entropy and velocity in a unified framework.

The observation of a system by an observer in another system, requires the connection between those two systems and the mutual exchange of information. The underlying assumption of this postulate is that observers can connect and exchange information. Once a connection is made between two reference systems, an addition rule between the respective stationary velocities applies. The addition rule follows the properties of commutativity and associativity, that is, symmetry and transitivity, respectively, which is required for the connection among reference frames. Given the addition rule $V_{A \oplus B} = f(V_A; V_B)$, the velocity of B measured from A, $V_{B,A}$, and its inverse, the velocity of A measured from B, $V_{A,B}$, are connected with $0 = f(V_{A,B}; V_{B,A})$ (symmetry), while $V_{A,B}$ is connected with the velocities of A and B measured from O with $V_{A,O} = f(V_{A,B}, V_{B,O})$ (transitivity).

Consider two originally independent systems A and B, with velocities $V_A$ and $V_B$, respectively, as measured by a third system O. The addition of velocities requires the exchange of information through some connection, eventually reaching a stationary state, where the velocity of the whole system, $V_{A \oplus B}$ is also stationary. For now, we do not focus on any particular addition rule, i.e., this may be Galilean, relativistic, or even more broadly defined. In general, the addition rule provides the relationship $V_{A \oplus B} = f(V_A, V_B)$, where the 2-D symmetric function *f(x,y)* forms a mathematical group on the set of velocities: (i) Closure: For any two velocities $V_A$ and $V_B$ belonging to the set of possible velocities $\Omega_V$, their addition $V_{A+B}$ belongs also to $\Omega_V$; in fact $\Omega_V$ is bounded, since their measure, the speed, is given by $0 \leq V \leq c$, where *c* denotes the upper limit of speed values, the speed of light in vacuum, i.e., $0 \leq f(V_A, V_B) \leq c$. (ii)

Identity: if $V_B = 0$, then for any $V_A$, $V_{A+B} = V_A$, or $f(V_A,0) = V_A$. (iii) Inverse: for any $V_A$, there exists its inverse element with velocity $\bar{V}_A$, for which $f(V_A,\bar{V}_A) = 0$. (iv) Associativity: The velocity of B measured by A, $V_{B,A}$, is expressed by the addition function of the velocity of A measured by C, $V_{A,C}$, and the velocity of C measured by B, $V_{C,B}$, i.e., $V_{A,B} = f(V_{A,C}, V_{C,B})$. (v) Commutativity: The group is abelian, since the addition function is symmetric, $f(V_A,V_B) = f(V_B,V_A)$. (vi) Finally, the existence of a fixed and invariant speed $c$, requires $f(V_A,c) = c$ for any $V_A$.

We note two key things in this development. First, the symmetry of the general addition rule, $0 = f(V_{A,B}; V_{B,A})$ does not necessarily lead to $V_{B,A} = V_{A,B}$, as in the case of Einstein's special relativity, (indeed, if $(V+u)/(1+Vu/c^2)=0$, then $V=-u$). Still, they are mutually inverse values $V_{A,B} = \bar{V}_{B,A}$ or $V_{B,A} = \bar{V}_{A,B}$, but with the associated inverse element definition given above. Second, both the addition of entropies (*H*-partitioning) and the addition of velocities form a mathematical group on their allowable set of values [71].

Therefore, the first postulate is as follows: *There is no absolute reference frame in which speed is zero*. On the contrary, the velocities are connected with the properties of symmetry and transitivity, broadly defined under a general addition rule that forms a mathematical group on the set of allowable speeds.

### *4.2. Second Postulate: Existence of a fixed and invariant speed, upper limit of speed values*

Einstein's second postulate of relativity is that the speed of light is fixed (stationary, i.e., constant with time) and invariant (i.e., constant for all observers). Throughout this paper, we have this noted with *c*, however, the essence of this postulate is the existence of a nonzero, finite speed, fixed and invariant among any time and observers, and not specifically that *c* is the speed of light in a vacuum. This is because neither the nature of light nor the value of the certain fixed speed is involved in the formalism of relativity. Specifically, in the speed addition rule, there is no requirement that the involved fixed speed *c* refers to the speed of light, only that this speed *c* is fixed and invariant. Indeed, adding *u*=*c* on an arbitrary speed *V*, results in the same speed *c*. Nevertheless, such a fixed and invariant speed, if exists, is also the maximum speed; this is a consequence, and not requirement of the postulate.

Therefore, the only requirement of the postulate is the existence of a nonzero, finite speed, *c*, fixed and invariant in time and among all observers. Below, we explain these terms; (all the involved velocities are considered in the same direction, for simplicity).

- *Fixed velocity (constant for all times)*. A fixed velocity, $V_*$, means it is a stable fixed-point solution in the velocity addition rule. In order to explain the terms, let the addition rule between two arbitrary velocities, $V_A$ and $V_B$, $V_{A+B} = f(V_A,V_B)$; now consider the case of sequential additions of the speed fluctuation $\delta u_i$ added to the velocity $V_i$, with *i* numbering the iterated addition, $V_{i+1} = f(V_i, \delta u_i)$. The term "fixed-point" of the velocity $V_i$ means $V_* = f(V_*, \delta u_i)$, independently of the value of $\delta u_i$. Then, the term "stable" characterizes the type of stability of the fixed point. Stable fixed point means it "attracts" the iterated velocity, leading eventually to smaller deviations: $|V_{i+1} - V_*| < |V_i - V_*|$. Therefore, the stable fixed point can be approached but not reached, i.e., there is no finite iteration step $i=j$, for which $|V_j - V_*| = 0$. On the other hand, the stability allows the existence of elements with velocity equal to the fixed point at all the iterations (eternally), i.e., if for some $i=j$, there is $|V_j - V_*| = 0$, the same holds for all $i$'s, from $i=0$ and beyond. For instance, the addition rule in Einstein's special relativity gives the iterated velocity values $V_{i+1} = f(V_i, \delta u_i) = (V_i + \delta u_i)/(1 + V_i \delta u_i / c^2)$, thus, the stable fixed velocity is given by $V_* = f(V_*, \delta u_i) = (V_* + \delta u_i)/(1 + V_* \delta u_i / c^2)$, leading to $V_* = c$; as expected, the speed of light $c$ involved in the relativity addition rule is given by the fixed velocity.

  - *Invariant velocity (constant for all observers)*. To understand the invariance of velocities, we need first to set how different velocities can be connected (third postulate). The velocity can be measured by different observers, whose connection is subject to a rule of addition of velocities. In particular, the velocity of A measured by B, $V_{A,B}$, is a function of the velocity of A measured by C, $V_{A,C}$, and the velocity of C measured by B, $V_{C,B}$, i.e., $V_{A,B} = f(V_{A,C}, V_{C,B})$. Once we realize that the velocity connection is determined by an addition rule, the invariant velocity can be determined in a way similar to the fixed velocity. Namely, the velocity of A is the same for any observer, B or C, $V_{A,B} = V_{A,C}$; setting this as $V_{A,B} = V_{A,C} \equiv V_*$, we can find $V_*$ from $V_* = f(V_*, V_{C,B})$, which leads to the value of $V_*$, independently of the value of $V_{C,B}$. Again, Einstein's relativity leads to $V_* = c$, thus, as expected, the speed $c$ involved in the relativity addition rule is invariant for all the observers.

The velocity provides information for both speed and direction, and thus, the fixed and invariant velocity allows for different fixed and invariant speed values in the positive and negative directions, giving insights for the anisotropic adaptation of relativity; (to be examined in Section 5.2). The reasoning behind the existence of a fixed and invariant speed is the existence of a speed limit, and in particular, an upper limit, which can be approached but not reached. (The fixed and invariant speed constitutes an upper and not a lower limit, because the zero speed exists as a possibility.) Thus, a body with speed $V$ less than the fixed and invariant speed $c$, may approach, but never reach this limit speed, $V < c$; on the other hand, natural elements with this speed (e.g., photons in vacuum), will remain with this speed, eternally, $V = c$.

Therefore, we restate the second postulate as follows: *There exists a nonzero, finite speed, which remains fixed (i.e., constant for all times) and invariant (i.e., constant for all observers), thus it has the same value in all inertial frames of reference and constitutes the upper limit of any speed.*

### *4.3. Third Postulate: Existence of stationary velocities*

Stationarity is considered as a trivial condition in Einstein's special relativity and was not included explicitly as postulate. However, it is a requirement that restricts the generality of the addition rule that can apply to velocities.

If a system is stationary according to an inertial observer, it will be stationary for all inertial observers. In particular, if the velocity of A measured from observer B is stationary, i.e., $V_{A,B}$=constant, then, it would be stationary for any other stationary observer, e.g., an observer C with stationary velocity as measured from B, $V_{C,B}$=constant, i.e., C will also observe a stationary velocity for A, i.e., $V_{A,C}$=constant. This property of the existence of stationarity for all observers, leads to the *H*-partitioning or kappa addition described by Eq.(3).

Indeed, the generalized partitioning in Eq.(3) characterizes any physical quantity with the properties of entropy defect, i.e., symmetry, separability, and boundedness [1,2], or the existence of stationary entropy values [58,60]; therefore, it applies to both entropy and velocity. Then, the most general addition function *f* is described through the *H*-partitioning, that is,

$$V_{A \oplus B} \equiv V_A \oplus_\kappa V_B = H^{-1}\left[H(V_A) + H(V_B) - \tfrac{1}{\kappa} H(V_A) H(V_B)\right], \tag{10}$$

where the partitioning function *H* falls under the properties discussed in Section 2.

The question that arises now is whether the velocity addition of Einstein's special relativity can be carried out simply through the kappa addition of Eq.(10). Indeed, by selecting the partitioning function

$$H(V) = V / (1 + \tfrac{1}{2\kappa} V), \tag{11}$$

and substituting in Eq.(10), we end up with (see Supplementary, Section B.1):

$$V_{A \oplus B} \equiv V_A \oplus_\kappa V_B = (V_A + V_B) / [1 + \tfrac{1}{(2\kappa)^2} V_A V_B]. \tag{12}$$

Then, we derive kappa as a function of the corresponding invariant and fixed speed *c*. This can be found by setting $V_{A \oplus B} = V_A = c$ in the addition rule of Eq.(12), leading to

$$V < c = 2\kappa. \tag{13}$$

Therefore, Eq.(10) includes the standard velocity addition of Einstein's special relativity.

We can derive the corresponding extensive measure of velocity, $V_\infty$, following the formulation given by Eq.(5), and apply the specific partitioning function *H* given by Eq.(11). We find,

$$V_\infty = \ln[1 - \tfrac{1}{\kappa} H(V)]^{-\kappa} = -\tfrac{1}{2} \cdot 2\kappa \cdot \ln\left(\frac{1 - \tfrac{1}{2\kappa} V}{1 + \tfrac{1}{2\kappa} V}\right) = c \cdot \ln\left(\sqrt{\frac{1 + \tfrac{1}{c} V}{1 - \tfrac{1}{c} V}}\right), \tag{14a}$$

$$\text{or } \beta_\infty = \ln\left(\sqrt{\frac{1 + \beta}{1 - \beta}}\right), \text{ with } \beta \equiv V / c. \tag{14b}$$

We note that this is the so-called rapidity, a commonly used additive measure of velocity (e.g., [72,73]), but here we showed how can be naturally derived within the context of thermodynamic relativity.

The connection of rapidity with kinetic energy is identical with the relationship between extensive entropy and momentum, $\partial S_\infty / \partial U = 1/T$, as shown in Eqs.(7,8), i.e.,

$$\frac{\partial V_\infty}{\partial E} = \frac{1}{p} \ , \tag{15a}$$

where the standard energy momentum relativity equations have been used, $E = \gamma E_0$, $pc = \beta \gamma E_0$. (We mention again that, for simplicity, the paper takes the velocity and momentum as 1-dimensional, i.e., along the direction of motion.) This similarity between the extensive measures of entropy and velocity and their relationship with energy was inspiring for the development of thermodynamic relativity.

Following the symmetry and transitivity properties for both of entropies and velocities, we come to equations similar to Eqs.(8,9). Namely, between two connected systems A and B, we have $V_{A,B} \oplus V_{B,A} = 0$, or equivalently, $V_{\infty A,B} + V_{\infty B,A} = 0$, thus $\partial V_{\infty A,B} = -\partial V_{\infty B,A}$. Also, the conservation of energy requires $E_A + E_B = const.$, or $\partial E_A = -\partial E_B$, hence,

$$\frac{\partial V_{\infty A,B}}{\partial E_A} = \frac{\partial V_{\infty B,A}}{\partial E_B} = \frac{1}{p} \ , \tag{15b}$$

with a common momentum absolute value, given by $p = |p_{A,B}| = |p_{B,A}|$.

We note that variations of particles numbers $N$ and/or volume V are not part of the thermal equilibrium approach; surely, once $N$ and/or V are not fixed, then, they are involved in the first law of thermodynamics, i.e., $dU = TdS + \mu dN - PdV$. In the same way, variations of positions are not involved in the stationarity approach of relativistic kinematics, but once potential energy $\Phi$ is taken into account, then, the positions are involved in the energy equation, i.e., $dE = pdV + d\Phi(x)$. Thus, in order to describe the nature of kinematic and thermodynamic stationary states, there is no need for the size ($N$,V) and positional ($x$) variables to vary. Since we describe the common physical framework of entropy and velocity, we eliminate all of the other factors from the analysis and the energy variation and simply compare the two equations: $dU = TdS$ with $dE = pdV$, or $(\partial S_\infty / \partial U)_{size} = 1/T$ with $(\partial V_\infty / \partial E)_{size} = 1/p$.

Moreover, we must delineate the role of the third postulate compared to the first two: (1) The third postulate does not require the existence of an immobile reference frame; however, it also does not state whether such a frame actually exists or not; indeed, the classical perception of aether is to be immobile, i.e., the reference frame of zero velocity and entropy. It is the first postulate that states that this privileged frame does not exist. (2) The third postulate allows for the existence of a fixed and invariant speed; however, this does not mean that such a velocity necessarily exists. The second postulate states that this exists.

Therefore, we state the third postulate as follows: *If a system is stationary for a reference frame O, it will be stationary for all reference frames being stationary for O.*

**5. Relativity of entropy and velocity – A unified framework of thermodynamics and kinematics**

*5.1. Motivation*

It is remarkable that the relativity concepts for entropy and velocity are based on three identical postulates: (1) no privileged reference frame with zero value; (2) existence of an invariant and fixed value for all frames; and (3) existence of stationarity. Hereafter, we use the term reference frame to encompass both the kinematic reference frame of measuring another system's motion and the thermodynamic reference frame of measuring another system's entropy.

Any comparison of observations between reference frames requires a connection, through which information can be exchanged. We have shown that the *H*-partitioning via the kappa addition supplies the connection, which is characterized by the properties of symmetry and transitivity that underly the zeroth law of thermodynamics. Thus, postulates and *H*-partitioning (derived from the postulates) are all identical for both entropies and velocities, and hereafter we are using a common formulation, expressed in terms of the variable $x = S$ and $V$.

The kappa addition for entropies and velocities is given by the *H*-partitioning, shown in Eq.(3) and Eq.(10), for any well-defined function *H*, described by the properties set in Section 2; in both cases the kappa addition forms a mathematical group on their allowable set of values. Table 1 summarizes the characteristics of *H*-partitioning. Given a well-defined partitioning function *H*, we can formulate a kappa-addition rule that applies to entropies or velocities, and then, construct generalized schemes of nonlinear relativity. In this paper, we develop an anisotropic version of linear relativity (a.k.a., corresponding to a linear Lorentz transformation) as just one example.

**Table 1. Characteristics of *H*-partitioning (kappa-addition)**

| Property | Expression |
|---|---|
| 1. Partitioning Function | $H(x)$ |
| 2. Extensive Measure, $x_\infty$ | $\ln\left[1 - \frac{1}{\kappa} H(x)\right]^{-\kappa}$ |
| 3. Addition Rule $x_{A \oplus B} \equiv x_A \oplus_\kappa x_B$ | $H^{-1}\left[H(x_A) + H(x_B) - \frac{1}{\kappa} H(x_A) H(x_B)\right]$ |
| 4. Addition Rule, expressed as product | $1 - \frac{1}{\kappa} H(x_{A \oplus B}) = [1 - \frac{1}{\kappa} H(x_A)] \cdot [1 - \frac{1}{\kappa} H(x_B)]$ |
| 5. Subtraction $x_{A \oplus \bar{B}} \equiv x_A \oplus_\kappa \bar{x}_B$ | $H^{-1}\left[\dfrac{H(x_A) - H(x_B)}{1 - \frac{1}{\kappa} H(x_B)}\right]$ |
| 6. Inverse, $\bar{x}$ | $\bar{x} = H^{-1}\{-H(x) / [1 - \frac{1}{\kappa} H(x)]\}$ |

The partitioning function for entropies and velocities are $H(x) = x$ and $H(x) = x / (1 + \frac{1}{2\kappa} x)$, respectively; both cases can be written using a parameter, *a*, under the scheme $H_a(x) = x / (1 + \frac{a}{\kappa} x)$, where $a = 0$ characterizes entropies with the addition rule of the standard entropy defect (nonextensive thermodynamics), while $a = 1/2$ characterizes velocities with the relativistic addition rule (Einstein's special relativity).

We note that one may check that if we chose, instead of $H(x) = x$, its inverse (in terms of *κ*-addition), i.e., $H(x) = \bar{x} = -x / (1 - \frac{1}{\kappa} x)$, we would end up with the same addition rule, because $S_{A \oplus B} = S_A \oplus_\kappa S_B$ and $\bar{S}_{A \oplus B} = \bar{S}_A \oplus_\kappa \bar{S}_B$ are identical; this choice is not permissible (since $H<0$), but it is interesting that its functional form is similar to $H_a(x)$. We also recall that the double inverse returns the identity, i.e., $\bar{\bar{x}} = x$; this is true as long as the operations are characterized by the same kappa. It is interesting to consider the case where the second inverse operation may act with a different kappa.

The choice of focusing on the particular partitioning function $H(x) = H_a(x)$ has one additional important motivation. When it comes to relativity for velocities, $x=V$, the partitioning leads to the proportionality $1 - \frac{1}{\kappa}H(V') \propto [1 - \frac{1}{\kappa}H(V)]$ (Table 1, #4), where $V$ and $V'$ denote the velocities of a body measured in two different inertial reference frames. Then, the corresponding Lorentz transformation of spacetime coordinates is linear only when the relationship between the velocities is a rational linear function (at its greatest complexity), i.e., $V' = (a_1 V + a_2)/(a_3 V + a_4)$. This is true, only when $H$ is also a rational linear function. Also considering the properties of $H$-partitioning function (Section 2), we end up $H_a(x)$ with as the most general $H$-partitioning function aligned with linear Lorentz transformation.

*5.2. Anisotropic relativity*

We consider the case where the kappa is different in positive and negative entropies. Recall that negative entropy, $\bar{S}$, is defined as the quantity that when added to an entropy, returns zero, $S \oplus_\kappa \bar{S} = 0$; thus, if $S_{B,A}$ is the entropy of B measured by A, then its inverse, $S_{A,B}$, is the entropy of A measured by B.

Since there is no absolute entropic zero, and entropy is relative rather than absolute, there are no meaningful positive or negative absolute entropic values. Instead, there are positive or negative relative entropic values; namely, if $S_{A,B} > 0$, means that the entropy of A is larger than that of B, as measured by any other reference frame O, $S_{A,O} > S_{B,O}$. On the other hand, its inverse would be $S_{B,A} < 0$, i.e., again $S_{A,O} > S_{B,O}$, for any O. We define the "positive direction" to be that corresponding to $\Delta S > 0$, e.g., the observation of A from B to the previous example that corresponds to $S_{A,B} > 0$. Similarly, we call the "negative direction", its inverse, corresponding to $\Delta S < 0$, e.g., the observation of B from A to the previous example that corresponds to $S_{B,A} < 0$.

There is nothing different in entropy and velocity for the concepts of positive and negative directions, but perhaps are more easily understood for velocities, rather than entropies. For both the velocities and entropies, there is no "absolute direction" in the universe to be assigned as positive direction, or its inverse, as negative direction, independent of observers. In contrast, the concept of directions is itself relativistic, rather than absolute. When the distance of a body B, as measured from a reference frame A, increases, this defines the positive direction of velocity, with respect to A. At the same time, the distance of A increases with respect to the reference frame of B. The inverse direction defines the negative direction. In the negative direction with respect to A, the distance of B should decrease with a negative relative velocity, $\bar{V}_{B,A} < 0$; similarly, in the negative direction with respect to B, the distance of A should decrease with negative relative velocity, $\bar{V}_{A,B} < 0$. When the velocity $V_{B,A}$ is positive (or negative), the velocity of B is larger (or smaller) in magnitude from the velocity of A, as measured from any other observer O. The same characteristics identify the positive and negative directions for the entropic values. In particular, the entropy of B as measured from A can be positive assigning the positive direction, $S_{B,A} > 0$, or negative $\bar{S}_{B,A} < 0$ assigning the inverse, negative direction. When entropy $S_{B,A}$ is positive (negative), the entropy of B is larger (smaller) than the entropy of A, as measured from observer O. (See Figure 2.)

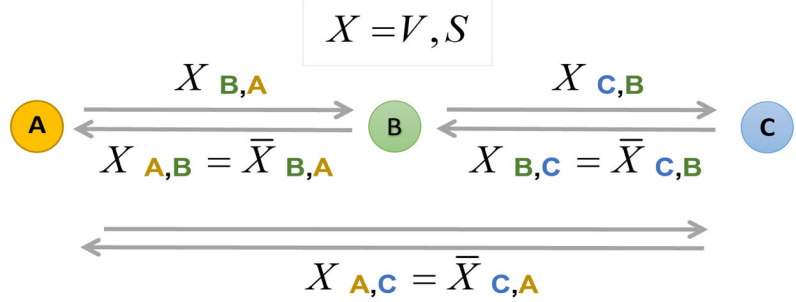

**Figure 2.** Relative positive and negative directions for velocities and entropies.

Next, we ask, what would be the upper limit of entropic values for the two directions? Given $H(x)$ and setting $0 = H(0) = H(x) + H(\bar{x}) - \frac{1}{\kappa}H(x)H(\bar{x})$, we find

$$H(\bar{x}) = -H(x)/[1 - \frac{1}{\kappa}H(x)], \tag{16}$$

from which we can extract the expression of the inverse, $\bar{x}(x) = H^{-1}\{-H(x)/[1 - \frac{1}{\kappa}H(x)]\}$. Then, we use this relation to find the upper limit of $|\bar{x}|$, given the upper limit of $x \leq H^{-1}(\kappa)$. We examine three cases, (1) Relativity for entropy, (2) Relativity for velocity, and (3) Relativity for both entropy and velocity.

*(1) Relativity for entropy, x=S, (nonextensive thermodynamics).*
The partitioning function is $H(x) = x$, thus $H(x) = x \leq \kappa$ or $x \leq \kappa$ is the upper limit in the positive direction. However, $H(\bar{x}) = \bar{x} = -x/(1 - \frac{1}{\kappa}x)$, leading to $|\bar{x}| = x/(1 - \frac{1}{\kappa}x) \leq +\infty$; namely, there is no upper limit in the negative direction.

*(2) Relativity for velocity, x=V, (= Einstein's special relativity).*
The partitioning function is $H(x) = x/(1 + \frac{1}{2\kappa}x)$, thus, $H(x) = x/(1 + \frac{1}{2\kappa}x) \leq \kappa$ or $x \leq 2\kappa$ provides the upper limit in the positive direction. Also, $H(\bar{x}) = \bar{x}/(1 + \frac{1}{2\kappa}\bar{x}) = -x/(1 - \frac{1}{2\kappa}x)$, hence, $\bar{x} = -x$, or $|\bar{x}| = x \leq 2\kappa$ provides the upper limit in the negative direction; namely, the upper limit in both directions is the same; Einstein's special relativity is isotropic [74].

*(3) Relativity for entropy and velocity, x=S or V, (a possible description of anisotropic relativity).*
We start with the partitioning function $H_a(x)$, that is,

$$H_a(x) = x/(1 + \frac{a}{\kappa}x), \tag{17}$$

which recovers the previous two cases for $a$=0 and $a$=1/2, respectively. Then, $H(x) = x/(1 + \frac{a}{\kappa}x) \leq \kappa$ or $x \leq \frac{1}{1-a}\kappa$ is the upper limit in the positive direction, while $H(\bar{x}) = \bar{x}/(1 + \frac{a}{\kappa}\bar{x}) = -x/(1 + \frac{a-1}{\kappa}x)$, $\bar{x} = -x/(1 - \frac{1-2a}{\kappa}x)$, or $|\bar{x}| = x/(1 - \frac{1-2a}{\kappa}x) \leq \frac{\kappa}{a}$ is the upper limit in the negative direction. Therefore, there are two kappa values, characterizing the upper limits of entropy and velocity in the positive and negative directions. Let $\kappa_1$ and $\kappa_2$ be these upper limits, respectively, i.e., for $0 < x$: $x < \kappa_1$, and for $\bar{x} < 0$: $|\bar{x}| < \kappa_2$. Then, the limits are equal to:

$$x < \kappa_1 \equiv \frac{1}{1-a}\kappa \text{ and } |\bar{x}| < \kappa_2 \equiv \frac{1}{a}\kappa. \tag{18}$$

The measurable value of upper limit should be given by the mean of the two directional limits. This defies Einstein's synchronization convention, which had assumed that the one-way speed is equal to the two-way speed; however, all experimental predictions of the theory do not depend on this convention (e.g., [75]). The inverse kappa measures the correlations [76,77], and thus, the mean value of the upper limits is given by the harmonic mean, i.e.,

$$\frac{1}{\kappa_O} \equiv \frac{1}{2}\left(\frac{1}{\kappa_1} + \frac{1}{\kappa_2}\right) = \frac{1}{2\kappa}, \text{ or } \kappa_O = 2\kappa. \qquad (19)$$

The anisotropy materializes the nonzero difference of these two upper limits, i.e.,

$$\frac{1}{\kappa_1} - \frac{1}{\kappa_2} = \frac{1-2a}{\kappa}, \qquad (20)$$

while the product of the two limits gives

$$\frac{1}{\kappa_1} \cdot \frac{1}{\kappa_2} = \frac{a(1-a)}{\kappa^2}. \qquad (21)$$

As shown in Table 1, the parameter $\kappa$ is included in the developed of this systematic formalism. Hereafter, the original parameterization of $\kappa$ and $a$ can be substituted by the directional kappa parameters of $\kappa_1$ and $\kappa_2$, as shown in Eq.(19) (for solving in terms of $\kappa$) and Eq.(20) (for solving in terms of $a$). For instance, the particular partitioning function $H_a(x)$ in Eq.(17) is now expressed as

$$H(x) = x/(1 + \tfrac{1}{\kappa_2}x), \qquad (22)$$

while the extensive measure $1 - \tfrac{1}{\kappa}H(x) = e^{-\tfrac{1}{\kappa}\cdot x_\infty}$ depends on both directional kappa,

$$1 - \tfrac{1}{\kappa}H(x) = (1 - \tfrac{1}{\kappa_1}x)/(1 + \tfrac{1}{\kappa_2}x). \qquad (23)$$

We recall that entropy $S$ and velocity $V$ can be noted using a common symbol, $x$. and use this symbol in the section. The limiting symbols of $\{\kappa_1, \kappa_2, \kappa_O\}$ refer to both entropy and velocity, unless we focus specifically on the velocity relativity and use $\{\kappa_1, \kappa_2, \kappa_O\} \to \{c_1, c_2, c_O\}$.

### 5.3. Anisotropic kappa addition

The kappa addition describes equivalently both entropies or velocities, for any partitioning function $H$. Here, we apply the specific partitioning function $H_a(x)$ (Eq.(17)) in the kappa addition (Table 1), i.e., $H_a(x_{A\oplus B}) = H_a(x_A) + H_a(x_B) - \tfrac{1}{\kappa}H_a(x_A)H_a(x_B)$, which after some calculus leads to the addition rule (see Supplementary, Section B.1):

$$x_{A\oplus B} \equiv x_A \oplus_\kappa x_B = \frac{x_A + x_B - \tfrac{1-2a}{\kappa}x_A x_B}{1 + \tfrac{a(1-a)}{\kappa^2}x_A x_B}, \qquad (24)$$

or, in terms of the directional upper limits:

$$x_{A\oplus B} \equiv x_A \oplus_\kappa x_B = \frac{x_A + x_B - (\tfrac{1}{\kappa_1} - \tfrac{1}{\kappa_2})x_A x_B}{1 + \tfrac{1}{\kappa_1\kappa_2}x_A x_B}. \qquad (25)$$

Notice the two limiting cases of standard entropy defect ($a=0$, maximum anisotropy) for $\kappa_2 \to \infty$, and special relativity ($a=1/2$, zero anisotropy), for $\kappa_1 = \kappa_2$.

The entropies or velocities may be expressed as normalized to the average upper limit, $\chi = S/\kappa_O$ or $\chi = V/c_O$; then, the kappa addition is written as

$$\chi_{A\oplus B} \equiv \chi_A \oplus \chi_B = \frac{\chi_A + \chi_B - r\chi_A\chi_B}{1+(1-\frac{1}{4}r^2)\chi_A\chi_B}, \tag{26}$$

with

$$r \equiv (\tfrac{1}{\kappa_1}-\tfrac{1}{\kappa_2})\cdot\kappa_O = 2(1-2a), \text{ and} \tag{27a}$$

$$\kappa_1 = \kappa_O/(1+\tfrac{1}{2}r) \text{ and } \kappa_2 = \kappa_O/(1-\tfrac{1}{2}r). \tag{27b}$$

### 5.4. Regular and anomalous anisotropy - Example of constant "acceleration"

We discuss the notion of regular $\kappa_1 < \kappa_2$ and anomalous $\kappa_1 > \kappa_2$ anisotropy. As an example, we examine the case of constant rate of increase of $x = S,V$, where both entropy and velocity are covered by $x$. For simplicity, we call the rate of change "acceleration", $dx/dt = a = $ const. (for both entropy and velocity).

For a continuous addition of entropy, $\Delta x = \Delta\sigma$, or velocity, $\Delta x = \Delta u$, in a time-scale of $\Delta t$, we construct the difference equation $x_n = f(x_{n-1})$ that connects the entropy or velocity $x_n$ of the $n^{th}$ iteration with the previous one $x_{n-1}$ of the $(n–1)^{th}$ iteration (e.g., [35,78]),

$$x_n = \frac{\Delta x + x_{n-1}\cdot[1-(\tfrac{1}{\kappa_1}-\tfrac{1}{\kappa_2})\Delta x]}{1+\tfrac{1}{\kappa_1\kappa_2}\Delta x\cdot x_{n-1}}. \tag{28a}$$

In the classical case of unrestricted addition, the respective equation for the value of $x$ is trivial: $x_n = x_{n-1}+\Delta x$, that is, solved to $x_n = x_0 + \Delta x \cdot n$. We also assume no initial entropy or velocity, for simplicity, i.e., $x_n = \Delta x \cdot n$. Time is measured by the number of iterations, i.e., $t = n\cdot\Delta t$, while the constant rate is $a = \Delta x/\Delta t$, hence, $x_n = a\cdot t$. In the negative direction, the equation is $\bar{x}_n = -a\cdot t$, and since $\bar{x} = -x$, the absolute value behaves similarly to the positive direction, $|\bar{x}_n| = a\cdot t$. Clearly, there is no upper limit on the values of $x$ in either direction, namely, entropy and velocity are unbounded in the classical case, $S_{t\to\infty} \to \infty$ and $V_{t\to\infty} \to \infty$.

Next, we repeat the previous steps, for the restricted addition of entropies or velocities, given by Eq.(28). In order to solve this difference equation, it is easier to use the corresponding $H$-partitioning function, Eqs.(22,23), and the formalism of the kappa addition (Table 1). We induce the recursive relation:

$$1-\tfrac{1}{\kappa}H(x_n) = [1-\tfrac{1}{\kappa}H(x_{n-1})]\cdot[1-\tfrac{1}{\kappa}H(\Delta x)] = \cdots = [1-\tfrac{1}{\kappa}H(x_0)]\cdot[1-\tfrac{1}{\kappa}H(\Delta x)]^n. \tag{28b}$$

Again, we assume zero initial values, and given $H(x_0 = 0) = 0$ (properties of function $H$, Section 2), we find that the solution is given by $1-\tfrac{1}{\kappa}H(x_n) = [1-\tfrac{1}{\kappa}H(\Delta x)]^n$. The partitioning function is $H(x_i) = x_i/(1+\tfrac{1}{\kappa_2}x_i)$ and $H(\Delta x) \cong \Delta x$ (for small $\Delta x$). Hence, in the limit of many iterations, we have $(1-\tfrac{1}{\kappa}\Delta x)^n = (1-\tfrac{1}{\kappa}\cdot\tfrac{x_\infty}{n})^n \xrightarrow{n\to\infty} e^{-\tfrac{1}{\kappa}\cdot x_\infty}$, where we substituted $\Delta x$ with the extensive measure of $x$, $x_{\infty n} = \Delta x\cdot n$ (that is, $x_n(\kappa\to\infty)$). Again, we introduce continuous time $t = n\cdot\Delta t$: $x_\infty = \Delta x\cdot n = (\Delta x/\Delta t)\cdot t = \alpha\cdot t$, thus, we write $x_n = x_t$; hence, we have $1-\tfrac{1}{\kappa}H(x_n) \cong e^{-\tfrac{1}{\kappa}\cdot x_\infty}$, or $1-\tfrac{1}{\kappa}H(x_t) = e^{-\tfrac{1}{\kappa}\cdot a\cdot t}$, where the equations become exact in the infinitesimal limits ($\Delta x \to dx$, $\Delta t \to dt$, with $a = dx/dt$). Substituting from Eq.(22b), $1-\tfrac{1}{\kappa}H(x_t) = (1-\tfrac{1}{\kappa_1}x_t)/(1+\tfrac{1}{\kappa_2}x_t)$, we solve in terms of $x_t$ and its inverse, $\bar{x}_t = -x_t/[1-(\tfrac{1}{\kappa_1}-\tfrac{1}{\kappa_2})x_t]$, i.e.,

$$x_t = \frac{1-e^{-\tfrac{1}{\kappa}\cdot a\cdot t}}{\tfrac{1}{\kappa_1}+\tfrac{1}{\kappa_2}e^{-\tfrac{1}{\kappa}\cdot a\cdot t}}, \quad \bar{x}_t = -\frac{1-e^{-\tfrac{1}{\kappa}\cdot a\cdot t}}{\tfrac{1}{\kappa_2}+\tfrac{1}{\kappa_1}e^{-\tfrac{1}{\kappa}\cdot a\cdot t}}, \tag{29}$$

with limits $x_\infty \to \kappa_1$ and $|\bar{x}_\infty| \to \kappa_2$.

Figure 3 plots the difference equations in Eq.(29) showing the isotropic case (b) and the two extreme cases of anisotropy: regular anisotropy, which is aligned with the concept of entropy defect (a) and the opposite extreme of anomalous anisotropy that does not align with the entropy defect (c). The entropy defect, which leads to a negative change on the system's entropy, is not due to the specific physical quantity involved, e.g., the thermodynamic meaning of entropy, but due to the "Cause-Effect" principle [2]. According to this, the insertion of some entropy $S_{in}$ into the system provokes a negative feedback of $-S_D$ (the entropy defect), and thus the total change of the system's entropy is $\Delta S = S_{in} - S_D \geq 0$, that is, positive, because $S_D \leq S_{in}$, i.e., the effect, $S_D$, is less than the cause, $S_{in}$. The cause-effect principle applies to both entropy and velocity, and thus, thermodynamic relativity has anisotropy: $\kappa_1 \leq \kappa_2$, while the anomalous case of $\kappa_2 \leq \kappa_1$ violates the "cause-effect" principle. Therefore, the two physically accepted extrema are: (i) $\kappa_1 = \kappa_2$ (e.g., the case of Einstein's special relativity for velocities) and (ii) $\kappa_1 < \infty$ with $\kappa_2 \to \infty$ (e.g., the case of the standard defect for entropies).

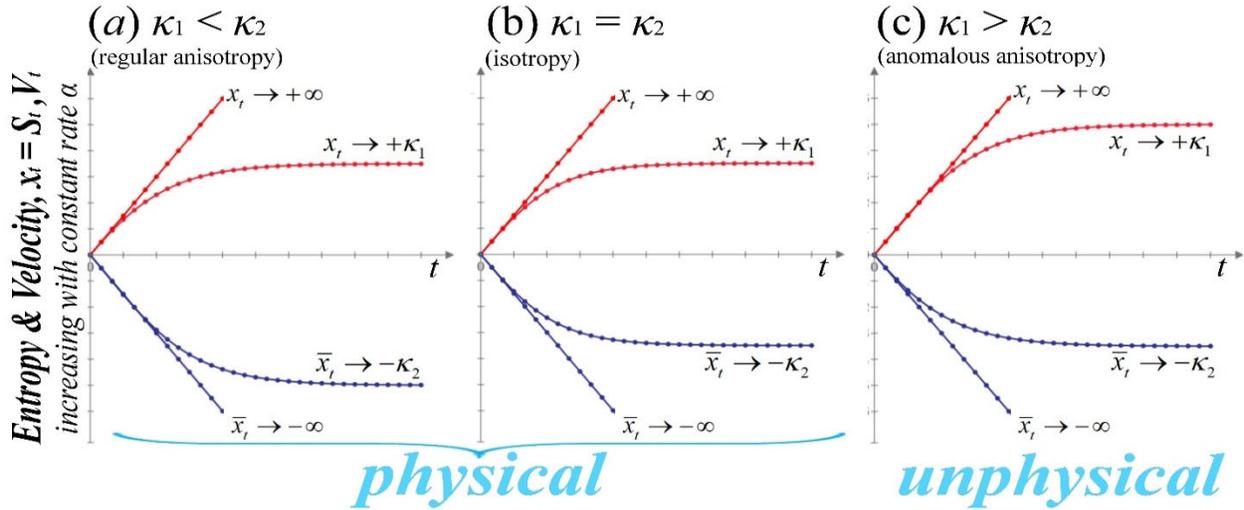

**Figure 3.** Entropy or velocity, $x_t = S_t$, $V_t$, increasing with time, $t=n\cdot\Delta t$, at a constant rate $\alpha=\Delta x/\Delta t$, plotted for finite limits $\kappa_1$ and $\kappa_2$ of the positive ($x_t$) and negative ($\bar{x}_t$) directions, respectively, and for the cases: (a) $\kappa_1 < \kappa_2$, (b) $\kappa_1 = \kappa_2$, and (c) $\kappa_1 > \kappa_2$ (anomalous anisotropy); $n$ counts the iterations, where each iteration has a time scale of $\Delta t$. Also shown are the cases of classical physics, where the limits are taken to infinity, $\kappa_1 \to \infty$, $\kappa_2 \to \infty$, the values of entropy or velocity continuously increase, unboundedly, towards infinity.

In the classical understanding, the entropy and velocity are allowed to constantly increase toward infinity. Einstein's special relativity restricts the velocities to increase up to the limit of light speed value, isotropically, i.e., for both the positive and negative directions. On the other hand, thermodynamics with simple entropy defect allows the entropy to increase up to a limit only in the positive direction, while it is unrestricted in the negative direction. The developed thermodynamic relativity naturally merges the two restrictions in a generalized conception of anisotropic relativity, where the positive and negative directions are characterized by different limits.

## 5.5. Comparison between the extreme cases and a natural generalization

The relativity of velocities and entropies is based on identical frameworks with no mathematical differences other than the quantities themselves. They are characterized by the same three postulates discussed in Sections 3 and 4 for entropies and velocities, respectively. Both frameworks are characterized by the *H*-partitioning, that is, the composition of the total value as a function of the constituents' values (entropies or velocities), which is expressed by the kappa addition. The partitioning characterizing Einstein's special relativity, $H(x)=x/[(1+x/(2\kappa)]$, is isotropic, namely, it considers equal finite fixed and invariant upper limits for the positive and negative directions, $\kappa_1=\kappa_2<\infty$. At the other extreme, the partitioning with entropy defect (nonextensive thermodynamics), $H(x)=x$, is the most anisotropic case possible, because while it has a finite fixed and invariant upper limit in the positive direction, $\kappa_1<\infty$, it has an infinite upper limit in the negative direction, $\kappa_2\to\infty$. The general case including these two extrema has a partitioning function that depends on any value of $\kappa_2$ (finite or not), $H(x)=x/(1+x/\kappa_2)$, as shown in Eq.(22). While other *H*-partitioning functions (following the properties presented in Section 2) may be suitable for constructing a relativity framework, the obvious generalization is the consideration of any finite upper limit in the negative direction. Table 2 summarizes the basic characteristics of relativity for entropies and velocities, for the disciplines of Einstein's special relativity and nonextensive thermodynamics, and their anisotropic generalization – thermodynamic relativity.

**Table 2. Comparison between relativity of velocities, entropies, and their common framework**

| Characteristic Property | Relativity for Velocities | Relativity for Entropies | Relativity for Velocities & Entropies |
|---|---|---|---|
| Physical Framework | Einstein's Special Relativity | Nonextensive Thermodynamics | Thermodynamic Relativity |
| Variable | $x = V$, $\partial x_\infty / \partial E = 1/p$ | $x = S$, $\partial x_\infty / \partial E = 1/T$ | $x = V$ or $S$ $\partial x_\infty / \partial E = 1/p$ or $1/T$ |
| Anisotropy | $a = 1/2$ | $a = 0$ | $1/2 \geq a \geq 0$, $a = \kappa_1/(\kappa_1+\kappa_2)$ |
| | $r = 0$ | $r = 2$ | $0 \leq r \leq 2$, $r \equiv (\frac{1}{\kappa_1}-\frac{1}{\kappa_2})\cdot\kappa_O = 2(1-2a)$ |
| | $R = 0$ | $R = 1$ | $0 \leq R \leq 1$, $R \equiv 1-\frac{\kappa_1}{\kappa_2} = \frac{1-2a}{1-a}$ |
| Partitioning Function $H$ | $H(x) = x/(1+\frac{1}{2\kappa}x)$ | $H(x) = x$ | $H(x) = x/(1+\frac{1}{\kappa_2}x)$ |
| Upper limits $x \leq \kappa_1$, $|\bar{x}| \leq \kappa_2$ | $\kappa_1 = \kappa_O$ $\kappa_2 = \kappa_O$ | $\kappa_1 = \frac{1}{2}\kappa_O$ $\kappa_2 \to \infty$ | $\kappa_1 = \kappa_O/(1+\frac{1}{2}r)$ $\kappa_2 = \kappa_O/(1-\frac{1}{2}r)$ |
| Mean limit (Observed) | $\frac{1}{\kappa_O} = \frac{1}{2}(\frac{1}{\kappa_1}+\frac{1}{\kappa_2}) = \frac{1}{2\kappa}$, $\kappa_O = 2\kappa$ | | |
| Extensive Measure, $x_\infty$ | (Rapidity) | (BG entropy) $-\kappa_1\cdot\ln(1-\frac{1}{\kappa_1}x)$ | $\kappa_O\cdot\ln\left(\sqrt{\frac{1+\frac{1}{\kappa_2}x}{1-\frac{1}{\kappa_1}x}}\right)$ |
| | $\kappa_{\mathrm{O}} \cdot \ln\left(\sqrt{\dfrac{1+\frac{1}{\kappa_{\mathrm{O}}}x}{1-\frac{1}{\kappa_{\mathrm{O}}}x}}\right)$ | | |
| Addition Rule $x_{\mathrm{A \oplus B}} \equiv x_{\mathrm{A}} \oplus_\kappa x_{\mathrm{B}}$ | $\dfrac{x_{\mathrm{A}}+x_{\mathrm{B}}}{1+\frac{1}{\kappa_{\mathrm{O}}^2}x_{\mathrm{A}}x_{\mathrm{B}}}$ | $x_{\mathrm{A}}+x_{\mathrm{B}}-\frac{1}{\kappa_1}x_{\mathrm{A}}x_{\mathrm{B}}$ | $\dfrac{x_{\mathrm{A}}+x_{\mathrm{B}}-(\frac{1}{\kappa_1}-\frac{1}{\kappa_2})x_{\mathrm{A}}x_{\mathrm{B}}}{1+\frac{1}{\kappa_1\kappa_2}x_{\mathrm{A}}x_{\mathrm{B}}}$ |
| Subtraction $x_{\mathrm{A \ominus \bar{B}}} \equiv x_{\mathrm{A}} \oplus_\kappa \bar{x}_{\mathrm{B}}$ | $\dfrac{x_{\mathrm{A}}-x_{\mathrm{B}}}{1-\frac{1}{\kappa_{\mathrm{O}}^2}x_{\mathrm{A}}x_{\mathrm{B}}}$ | $\dfrac{x_{\mathrm{A}}-x_{\mathrm{B}}}{1-\frac{1}{\kappa_1}x_{\mathrm{B}}}$ | $\dfrac{x_{\mathrm{A}}-x_{\mathrm{B}}}{1-(\frac{1}{\kappa_1}-\frac{1}{\kappa_2})x_{\mathrm{B}}-\frac{1}{\kappa_1\kappa_2}x_{\mathrm{A}}x_{\mathrm{B}}}$ |
| Inverse, $\bar{x}$ | $-x$ | $\dfrac{-x}{1-\frac{1}{\kappa_1}x}$ | $\dfrac{-x}{1-(\frac{1}{\kappa_1}-\frac{1}{\kappa_2})x}$ |
| Average, $x_0$ $x_0^{-1}=\frac{1}{2}(x^{-1}+\left|\bar{x}\right|^{-1})$ | $x$ | $\dfrac{x}{1-\frac{1}{2\kappa_1}x}$ | $\dfrac{x}{1-\frac{1}{2}(\frac{1}{\kappa_1}-\frac{1}{\kappa_2})x}$ |

Note: $E$ denotes, either the mechanical energy per particle in the relativity of velocities, or the internal energy $U$ per particle, that is, the average mechanical energy, in the relativity of entropies.

## 6. Formulation of thermodynamic relativity for velocity
### 6.1. Scheme of the derivations of thermodynamic relativity

The Lorentz transformation has played a central role in special relativity. It is a set of linear equations that bind space and time of two systems with a constant relative velocity between them. Originally, the transformation was derived as a mitigation of the apparent inconsistency between the constancy of the speed of light and the existence of aether. In particular, the negative result of Michelson–Morley's experiment [79] to determine the Earth's movement through the aether, suggested a unique explanation, the constancy of the speed of light, independent of the observer's motion through the aether. Then, Fitzgerald [80] suggested the theory that objects change length due to this type of movement, while Lorentz [81] independently presented the same idea in more mathematical detail. In parallel with Lorentz, Larmor [82] published an approximation to the Lorentz transformations. While all of these interpreted the Lorentz transformation in favor of the existence of aether, Einstein derived and interpreted the same formalism within the framework of relativity, abandoning the concept of aether's privileged frame.

The defining property of the Lorentz transformation is that it preserves the spacetime interval between any two events. Even though Einstein's postulates, the principle of relativity and the constancy of the speed of light for every observer, led to the derivation of the transformation, the aforementioned property actually provides the strict definition of the Lorentz transformation. Given the Minkowsky metric, a matrix broadly denoted with $\eta$, which models the flat spacetime of special relativity and is used for determining the "distance" between two spacetime events, the Lorentz transformation can be determined though the metric $\eta$. In this picture, the metric stands as another way of deriving the transformation, while the addition rule of velocities is just an unavoidable and useful outcome.

This paper follows a scheme similar to Einstein's relativity, starting from the relativity postulates and ending with the Lorentz transformation and its connection with Minkowsky metric, but in an opposite direction through the steps (Figure 4). Starting from the postulates (P), the addition of velocities (V) is not an outcome, but a necessary step, required for the derivation of the Lorentz transformation (L). The relationship between Lorentz transformation (L) and Minkowsky metric (M) is the same (this is Eq.(39), independent of what is given and what is derived), but the transformation is used to determine the metric, rather than vice-versa. The postulates of thermodynamic relativity lead to the surprising result of an asymmetric Lorentz transformation and a non-diagonal metric. These characteristics define the anisotropy of our suggested relativity theory.

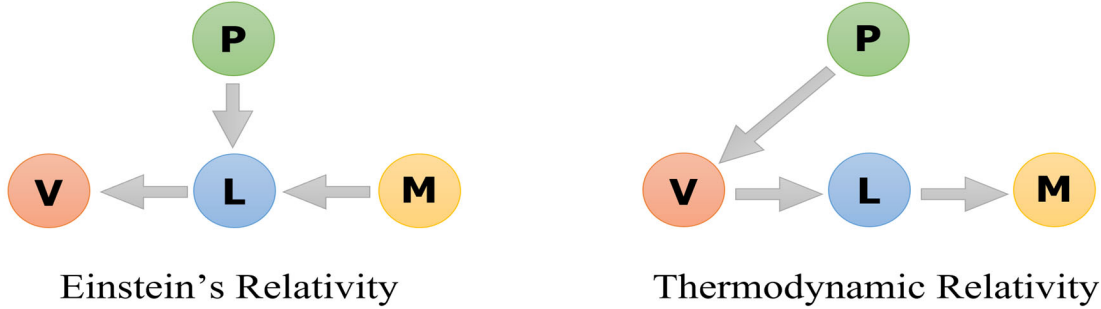

Einstein's Relativity                     Thermodynamic Relativity

**Figure 4.** Steps for the basic formulations of relativity: postulates (P), velocity addition (V), Lorentz transformation (L), and Minkowsky metric (M).

We continue with thermodynamic relativity of velocities – kinematics. We focus on the anisotropic relativistic kinematics, including anisotropy in the speed of light and velocity addition, asymmetric Lorentz transformation and non-diagonal metric, and finally, energy-momentum and energy-speed relations. Note that thermodynamic relativity should not be confused with a relativistic adaptation of thermodynamics; on the other hand, this theory can also shed light on the relativistic thermodynamics of continuous media (e.g., [83,84]).

### 6.2. Anisotropic speed of light

The speed of light (speed upper limit) in the positive and negative directions is respectively given by

$$c_1 = c_O / (1 + \tfrac{1}{2}r) \text{ and } c_2 = c_O / (1 - \tfrac{1}{2}r), \tag{30}$$

where the anisotropy is determined by

$$r \equiv \left(c_1^{-1} - c_2^{-1}\right) \cdot c_O. \tag{31}$$

This is consistent with the notation of previous anisotropic relativity concepts [75,85,86]. The observable speed of light is the harmonic mean:

$$c_O^{-1} \equiv \tfrac{1}{2}\left(c_1^{-1} + c_2^{-1}\right), \tag{32}$$

and defines the value of kappa, $\kappa = \tfrac{1}{2}c_O$, that characterizes the kappa addition. Indeed, to cover a distance $L$ light needs time $\tau_1 = L / c_1^{-1}$ and $\tau_2 = L / c_2^{-1}$, or, on average, $\tau_O \equiv \tfrac{1}{2}(\tau_1 + \tau_2)$, leading to Eq.(32).

The extreme cases are the following: (i) as $\tfrac{1}{2}r$ approaches 1, or $c_2 \to \infty$, light tends to propagate in negative direction instantaneously, while it takes the entire round-trip time to travel in the positive direction; (ii) As $\tfrac{1}{2}r$ tends to zero, both directions are characterized by the same light speed, $c_1=c_2=c_O$, formulating the isotropic case of Einstein's special relativity; and finally, (iii) for negative values of *r*, including the extreme case of $\frac{1}{2}r$ approaching -1, or $c_1 \to \infty$, we obtain the anomalous anisotropy, corresponding to violation of the "cause-effect" principle, as discussed in Section 5.5 and shown in Figure 3(c)).

*6.3. Velocity addition*

We consider one-dimensional velocities in the positive *V* and negative $\bar{V}$ directions. If a system B moves with respect to the system A with a velocity $V_B$, and A moves with respect to an observer O with a velocity $V_A$, then, B has a velocity measured from O given by the kappa-addition of velocities, i.e.,

$$V_{A \oplus B} \equiv V_A \oplus_\kappa V_B = \frac{V_A + V_B - (\frac{1}{c_1} - \frac{1}{c_2})V_A V_B}{1 + \frac{1}{c_1 c_2}V_A V_B} \text{ or } \beta_{A \oplus B} \equiv \beta_A \oplus \beta_B = \frac{\beta_A + \beta_B - r\beta_A \beta_B}{1 + (1 - \frac{1}{4}r^2)\beta_A \beta_B}, \quad (33)$$

where we used the notions of normalized velocities $\beta \equiv V/c_O$, and the anisotropy *r*.

The negative velocity $\bar{V}$ is expressed in terms of *V* as

$$\bar{V} = \frac{-V}{1 - (\frac{1}{c_1} - \frac{1}{c_2})V} \text{ or } \bar{\beta} = \frac{-\beta}{1 - r\beta}, \quad (34)$$

with $(1 - r\bar{\beta})(1 - r\beta) = 1$. This recovers the classical algebraic inverse $\bar{V} = -V$ when $c_1, c_2 \to \infty$.

Since the two magnitudes, *V* and $|\bar{V}|$, are not equal, we define the average directional velocity $V_O$; i.e., $V_O^{-1} \equiv \frac{1}{2}(V^{-1} + |\bar{V}|^{-1})$, following the same way of averaging as for the light speeds in Eq.(32). This gives:

$$V_O = \frac{V}{1 - \frac{1}{2}(\frac{1}{c_1} - \frac{1}{c_2})V} \text{ or } \beta_O = \frac{\beta}{1 - \frac{1}{2}r\beta}. \quad (35)$$

The two directions are limited by $V < c_1$ and $|\bar{V}| < c_2$, with the average velocity limited by $V_O < c_O$. The average velocity can be used when information is exchanged between two reference frames, such as in time-dilation (Section 7).

*6.4. Anisotropic Lorentz transformation*

Let a body move at velocity *V*, as measured from a reference frame O. If this frame O moves with a velocity *u* as measured from another reference frame O´, then, the velocity *V*´, as measured by O´, is given by the kappa addition of velocities:

$$V' = \frac{V(1 - r\frac{1}{c_O}u) + u}{1 + (1 - \frac{1}{4}r^2)\frac{1}{c_O^2}uV}. \quad (36)$$

Setting $V = dx/dt$, $V' = dx'/dt'$, we derive the linear Lorentz transformation (Supplementary, Section B.2),

$$\begin{pmatrix} c_O t' \\ x' \end{pmatrix} = L_r(\beta) \cdot \begin{pmatrix} c_O t \\ x \end{pmatrix}, \quad L_r(\beta) \equiv \gamma_r \cdot \begin{pmatrix} 1 & (1 - \frac{1}{4}r^2)\beta \\ \beta & 1 - r\beta \end{pmatrix}, \quad (37)$$

with the involved γ-factor, given by:

$$\gamma_r = 1/\sqrt{(1 - u/c_1) \cdot (1 + u/c_2)} \text{ or } \gamma_r = 1/\sqrt{(1 - \frac{1}{2}r\beta)^2 - \beta^2}. \quad (38)$$

*6.5. Non-diagonal metric*

The metric $\eta_r$ can be derived from the Lorentz transformation defining property that keeps the spacetime length $ds^2$ invariant. The Minkowsky metric, a matrix broadly denoted with $\eta_r$, models spacetime in special relativity and is used to derive the spacetime distance between any two events. The infinitesimal spacetime distance, $ds^2 = d\vec{x}^t \cdot \eta_r \cdot d\vec{x}$, remains invariant under the Lorentz transformation. The invariancy leads to the relationship between the metric and the Lorentz transformation and

$$L_r^t(\beta) \cdot \eta_r \cdot L_r(\beta) = \eta_r, \qquad (39)$$

(superscript $t$ denotes the transpose of the matrix of the Lorentz transformation). Using Eq.(39), we derive the form of the metric that corresponds to the transformation of Eq.(37) (Supplementary, Section B.4), i.e.,

$$\eta_r = \begin{pmatrix} -1 & \tfrac{1}{2}r \\ \tfrac{1}{2}r & 1-\tfrac{1}{4}r^2 \end{pmatrix}, \qquad (40)$$

reducing the Minkowsky metric, $\eta_0 = diag(-1,1)$, for $r=0$, as expected for recovering the isotropy.

### 6.6. Energy – Momentum equation

The 4-momentum vector (in one spatial dimension) is the energy-momentum expression $(E/c_\mathrm{O}, p)$ and provides the invariant quantity:

$$(E/c_\mathrm{O}, p) \cdot \begin{pmatrix} -1 & \tfrac{1}{2}r \\ \tfrac{1}{2}r & 1-\tfrac{1}{4}r^2 \end{pmatrix} \cdot \begin{pmatrix} E/c_\mathrm{O} \\ p \end{pmatrix} = -(E/c_\mathrm{O})^2 + rp(E/c_\mathrm{O}) + (1-\tfrac{1}{4}r^2)p^2. \qquad (41)$$

This is an invariant for any momentum $p$, thus, when taken for $p=0$, it equals to $-(E_0/c_\mathrm{O})^2$ that involves the energy $E_0$ at $p=0$, i.e.,

$$E^2 - r \cdot (pc_\mathrm{O}) \cdot E - \left[(1-\tfrac{1}{4}r^2)(pc_\mathrm{O})^2 + E_0^2\right] = 0. \qquad (42)$$

Solving in terms of energy,

$$E_\pm(p) = \tfrac{1}{2}r \cdot (pc_\mathrm{O}) \pm \sqrt{(pc_\mathrm{O})^2 + E_0^2}, \qquad (43a)$$

with $\pm$ corresponding to a positive and negative value. These two values are numerically equal for the isotropic case ($r=0$), but for the anisotropic case, we have:

$$|E_\pm| = \sqrt{(pc_\mathrm{O})^2 + E_0^2} \pm \tfrac{1}{2}r \cdot (pc_\mathrm{O}). \qquad (43b)$$

Equation (43a) or (43b) generalizes the isotropic standard energy-momentum form of Einstein's special relativity, $|E_\pm|_{r=0} = \sqrt{(pc_\mathrm{O})^2 + E_0^2}$, from which we recognize the matter (+) and antimatter (-) branches. The momentum is extensive (see Eq.(46b) below), thus, we may write $p_\pm = \pm p$. Then, Eq.(43b) can be written via a single function $G$, in a unified way for both matter and antimatter, such as,

$$|E_\pm| = E_0 \cdot G(p_\pm c_\mathrm{O} / E_0), \text{ with } G(x) = \sqrt{x^2 + 1} + \tfrac{1}{2}r \cdot x. \qquad (43c)$$

In Figure 5 we observe that for the matter branch, $E_+(p)$ is monotonically increasing, while for the antimatter branch, $|E_-|(p)$ has a local minimum at

$$p_{\min}c_\mathrm{O}/E_0 = \frac{\tfrac{1}{2}r}{\sqrt{1-\tfrac{1}{4}r^2}}, \quad |E_-|_{\min}/E_0 = \sqrt{1-\tfrac{1}{4}r^2}. \qquad (44)$$

The minimum shifts to infinity for $\tfrac{1}{2}r \to 1$, or equivalently, for $c_2 \to \infty$ (corresponding to maximum anisotropy, that is, the extreme case of nonextensive thermodynamics), leading to a monotonically decreasing function. On the contrary, the minimum shifts to zero for $\tfrac{1}{2}r \to 0$, or equivalently, for $c_2 = c_1$ (corresponding to isotropy, that is, the extreme case of standard relativity), leading to a monotonically increasing function. The minimum values of momentum and energy correspond to the smallest measurable values due to the presence of anisotropy; for isotropic relativity, these would both return to zero.

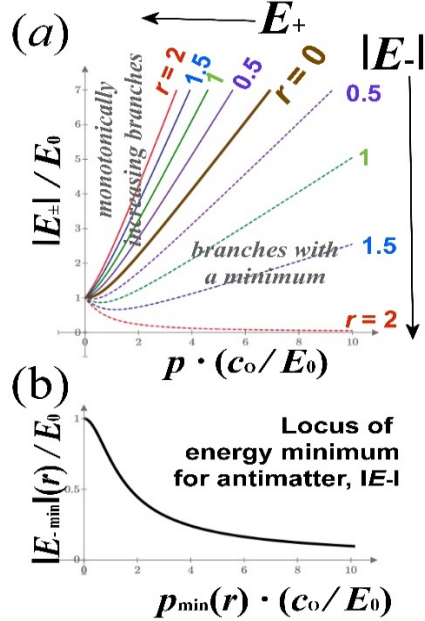

**Figure 5.** (a) Energy plotted as a function of momentum and for various values of the anisotropy $r$. Both the branches of matter, $E_+(p)$, and antimatter, $|E_-|(p)$, are shown. (b) Locus of momentum $p_{\min}(r)$ and energy $|E_-|_{\min}(r)$, corresponding to the energy minimum for the antimatter, plotted for all $0 \leq r \leq 2$.

### 6.7. Energy – Velocity equation

The momentum is given by

$$p = m\, dx'_{dx=0}/dt = \gamma_r \cdot \beta \cdot m c_O = m c_O \beta / \sqrt{(1-\tfrac{1}{2}r\beta)^2 - \beta^2}\ . \tag{45}$$

We note that the inverse momentum is

$$\bar{p} = \bar{\gamma}_r \cdot \bar{\beta} \cdot m c_O\ , \tag{46a}$$

where $\bar{\gamma}_r = \gamma_r \cdot (1 - r\beta)$ (Supplementary, Section B.3); given the identity $\bar{\gamma}_r \cdot \bar{\beta} = -\gamma_r \cdot \beta$, we find

$$p + \bar{p} = 0\ , \tag{46b}$$

that is, a result consistent with conservation of momentum.

Finally, we substitute Eq.(45) in Eq.(43b), and find:

$$|E_+| = \gamma_r E_0\ ,\ |E_-| = \bar{\gamma}_r E_0\ ,\ \text{that is,} \tag{47a}$$

$$|E_+| = E_0 / \sqrt{(1-\tfrac{1}{2}r\beta)^2 - \beta^2}\ ,\ |E_-| = E_0 \cdot (1 - r\beta) / \sqrt{(1-\tfrac{1}{2}r\beta)^2 - \beta^2}\ . \tag{47b}$$

## 7. Physical consequences of thermodynamic relativity

Here we explore several of the most famous physical consequences of special relativity, interwoven with conceptual or real experiments, in the context of thermodynamic relativity. These are (i) matter-antimatter baryonic asymmetry, (ii) time dilation, and (iii) Doppler effect. Emphasis is placed on the effect of anisotropy on the theoretical results and possible observations. The anisotropy $r$ may be impossible to be measured through purely kinematic effects, which are unavoidably connected to the problem of synchronisation [75]; here, we suggest ways of measuring the anisotropy to the extent that measurements can be completed out of the kinematics framework and the problem of synchronisation. The paradox of matter-antimatter asymmetry is examined first, in order to estimate an upper limit of the possible anisotropy that characterized the early universe. This anisotropy, if it existed, could have been spread throughout the evolved universe and thus might inform the magnitude of possible consequences on time dilation, the Doppler effect, and other future observations and tests.

### 7.1. Matter-antimatter asymmetry

The anisotropy described by thermodynamic relativity might impact the observed matter-antimatter baryon asymmetry in the universe. According to the formalism derived in Section 6.6 and Eq.(43b), the difference between the matter and antimatter energy branches (for the same $p$) is exactly proportional to anisotropy $r$,

$$E_+ - |E_-| = r \cdot (pc_O), \tag{48a}$$

where in the approximation of small speeds, we find $(E_+ - |E_-|)/E_0 = r \cdot (pc_O/E_0) = r \cdot \gamma_r(\beta)\beta \xrightarrow{\beta \ll 1} r \cdot \beta$.

The average speed $<\beta>$ of baryons is set to the thermal speed of the early universe, at the epoch of recombination. Matter decoupled from the cosmic background radiation about 300,000 years after the Big Bang, while the epoch of recombination occurred about 380,000 years after the Big Bang, where charged electrons and protons of the existed ionized plasma first formed hydrogen atoms,. A recombination temperature of ~4000 K (e.g., [87,88]) leads to a thermal speed $<\beta> \sim 3 \cdot 10^{-5}$, so $<E_+ - |E_-|> \sim 3 \cdot 10^{-5} r E_0$. The observed particle asymmetry is about one per billion of baryons (e.g., [89]), which can be interpreted in terms of energy anisotropy as $<E_+ - |E_-|> \sim 10^{-9} E_0$, corresponding to $r \sim 3 \cdot 10^{-5}$.

The smaller amount of energy deviation,

$$E_+(p_{min}) - |E_-|(p_{min}) = r \cdot p_{min}(r)c_O = E_0 \cdot \frac{\frac{1}{2}r^2}{\sqrt{1 - \frac{1}{4}r^2}}, \tag{48b}$$

where in the approximation of small anisotropies, we find $(E_+ - |E_-|)/E_0 \cong \frac{1}{2}r^2$; the previously mentioned matter-antimatter asymmetry, $<E_+ - |E_-|> \sim 10^{-9} E_0$, corresponds to a similar anisotropy, $r \sim 4.5 \cdot 10^{-5}$.

If such an anisotropy existed in the early universe, it could have spread throughout the evolved universe. Therefore, the anisotropy of $r \sim (4 \pm 1) \times 10^{-5}$ provides a possible starting point for modeling of other experiments and testing special relativity.

### 7.2. Time-dilation

Using the formalism of thermodynamic relativity we repeat Einstein's famous conceptual experiment of time-dilation for a light pulse [65]. The experiment involves two clocks consisting of a set of mirrors reflecting light back and forth. One of the clocks is fixed, while the other moves with constant velocity, relative to an inertial observer reading the clocks. (Note that the fixed clock is characterized as motionless, i.e., its position is constant with time, as opposed to the moving clock.) Then, according to the fixed clock, the one moving with constant velocity would be experiencing a time-dilation. Here we briefly describe the time-dilation within the framework of Einstein's special relativity, and then, proceed to thermodynamic relativity and the consequences if there is some level of anisotropy.

First, we derive the time-dilation from the Lorentz transformation, Eq.(37), set for a fixed clock ($\Delta x=0$):

$$\Delta t' = \gamma_r \cdot \Delta t = \Delta t / \sqrt{(1-\tfrac{1}{2}r\beta)^2 - \beta^2} \ . \tag{49a}$$

The observed dilated time depends on the average directional velocity, $V_O$ (see Eq.(35)); this velocity is used when information is exchanged between two reference frames. At a certain moment, $t=t'=0$, the two clocks are at the same position and synchronized. Then, at a distance $L$, the two clocks may exchange information on their readings, while the actual time-dilation is given by $\Delta t'_O$: $L = V \Delta t' = V_O \Delta t'_O$, or $\Delta t'_O = \Delta t' \cdot (V/V_O) = \Delta t' \cdot (1-\tfrac{1}{2}r\beta)$. Hence,

$$\Delta t'_O = \Delta t' \cdot (1-\tfrac{1}{2}r\beta) = \Delta t \cdot (1-\tfrac{1}{2}r\beta)/\sqrt{(1-\tfrac{1}{2}r\beta)^2 - \beta^2} = \Delta t/\sqrt{1-\beta_O^{\,2}} = \Delta t \cdot \gamma_O, \tag{49b}$$

where we observe that the formulation of time-dilation is similar to that of Einstein's special relativity, after substituting the isotropic $\gamma$-factor, $\Delta t'(r=0) = \Delta t \cdot \gamma_{r=0}$, to that of average velocity (see Eq.(46)), i.e., $\Delta t'_O = \Delta t \cdot \gamma_O$, with $\gamma_O = 1/\sqrt{1-\beta_O^{\,2}}$.

Alternatively, we derive the same result from the geometry between the pulses in the two reference frames. We consider the orthogonal triangle formed by the distance $L$ between the clocks, and the distances covered by the light pulse, according to the two reference frames, i.e., $\ell = c_O \Delta t$ and $\ell' = c_O \Delta t'_O$, for the fixed and moving frames, respectively. Using the Pythagorean theorem, $\ell'^2 = \ell^2 + L^2$ or $(c_O \Delta t'_O)^2 = (c_O \Delta t)^2 + (V_O \Delta t'_O)^2$ [90], we find again Eq.(49b), without the necessity of Lorentz transformation.

An early famous test of time-dilation concerned the decay of muons in their passage through the upper regions of the atmosphere. These muons are caused by collisions of cosmic rays with particles in the upper atmosphere, after which the muons reach Earth. If there were no time-dilation, then most of the muons should have decayed in the upper regions of the atmosphere, before reaching Earth. However, due to the time dilation of their lifetime, they are observed in considerably larger numbers. Measuring the number of the decayed muons, and given knowledge of the decay mean lifetime $\tau$, the total time travel in the atmosphere, and the average downward speed, provided a test for the validity of relativistic theory [91-94]. Given sufficiently precise measurements of the percentage of decayed muons and other parameters in the calculation, one might be able to detect an anisotropic relativity, $\exp[-t/(\gamma_O \tau)]$, which is larger than for the standard isotropic relativity. For example, given $\tau \sim 2.2$ μs [95] and for a time travel of $t \sim 5$ μs, and average speed of $\beta \sim 0.995$, an anisotropy $r \sim 10^{-5}$ would produce an additional ~0.02% muons reaching the lower atmosphere. While precision measurements of time-dilation may be possible generally, how precisely we know the parameters in the muon calculation may limit the testability through these specific observations.

### 7.3. Doppler effect

Another category of experimental tests is through the relativistic Doppler effect. Such an experiment was first performed by Ives & Stilwell [96], where they simultaneously observe a nearly longitudinal direct beam and its reflected image. Einstein [70,97] suggested the experiment based on the measurement of the relative frequencies. Similar experiments have been conducted with increasing precision [98-101].

We derive the relativistic Doppler effect on the wavelength of the emission propagating in the positive and negative directions, as follows. From the Lorentz transformation in Eq.(37), we have:

$$\Delta x' = \gamma_r \cdot \left[ u \cdot \Delta t + [1 - (\tfrac{1}{c_1} - \tfrac{1}{c_2})u]\Delta x \right] = \frac{u \cdot \Delta t + [1 - (\tfrac{1}{c_1} - \tfrac{1}{c_2})u]\Delta x}{\sqrt{(1 - u/c_1)\cdot(1 + u/c_2)}}. \tag{50}$$

In Eq.(50), we set $\Delta x \to \lambda$ and $\Delta x' \to \lambda'$, as well as $\Delta x = c_1 \Delta t$, and obtain the wavelength shifting

$$\lambda'(\beta) = \lambda \cdot \gamma_r \cdot (1 + \tfrac{1}{c_2}u) = \lambda \cdot \sqrt{\frac{1 + \tfrac{1}{c_2}u}{1 - \tfrac{1}{c_1}u}} = \lambda \cdot \sqrt{\frac{1 + (1 - \tfrac{1}{2}r)\beta}{1 - (1 + \tfrac{1}{2}r)\beta}}, \tag{51a}$$

which is a natural generalization of the respective quantity in Einstein's special relativity $\lambda' = \lambda \cdot \sqrt{(1+\beta)/(1-\beta)}$. The corresponding wavelength shifting for motion in the negative direction gives

$$\lambda'(\bar{\beta}) = \lambda \cdot \sqrt{\frac{1 + (1 - \tfrac{1}{2}r)\bar{\beta}}{1 - (1 + \tfrac{1}{2}r)\bar{\beta}}} = \lambda \cdot \sqrt{\frac{1 - (1 + \tfrac{1}{2}r)\beta}{1 + (1 - \tfrac{1}{2}r)\beta}}. \tag{51b}$$

Then, the combination of the effect on the wavelengths of the direct, $\lambda'_+ = \lambda'(\beta)$, and reflected, $\lambda'_- = \lambda'(\bar{\beta})$, emissions, is (in terms of wavelengths and/or frequencies):

$$\lambda'_+(\beta) \cdot \lambda'_-(\beta) = \lambda_0^2, \quad \nu'_+(\beta) \cdot \nu'_-(\beta) = \nu_0^2. \tag{52}$$

This relationship has already been verified with various experiments, some with a high precision (e.g., [99; 102-104]). Nonetheless, it cannot be used to test for anisotropy, as Eq.(52) is valid for any anisotropy $r$ or isotropy ($r$=0).

On the other hand, the average wavelength does depend on the anisotropy, and thus, it could be used for testing the theory; this is given by

$$\lambda'_O \equiv \tfrac{1}{2}[\lambda'_+(\beta) + \lambda'_-(\beta)] = \lambda \cdot \gamma_O. \tag{53}$$

The ratio $\lambda'_O / \lambda$ equals the ratio of time dilation, $\Delta t'_O / \Delta t$, thus, they have the same expansion in terms of $\beta$ and/or $r$, as shown in Eq.(49b), which can be used for testing anisotropic relativity.

### 7.4. Suggested experiment to measure r via the Doppler effect

We have seen that the formulations in terms of velocity for the time dilation, Eq.(49b), and Doppler effect, Eq.(53), is identical – both depend on the averaged gamma factor, $\gamma_O$. We also remark that their expression is independent of the choice of positive or negative directions, i.e., $\Delta t'_O(\bar{\beta}) = \Delta t'_O(\beta)$ and $\lambda'_O(\beta) = \lambda'_O(\bar{\beta})$. Indeed, as expected, the average velocity, $\beta_O$, has the same magnitude in the two directions, $\beta_O(\bar{\beta}) = -\beta_O(\beta)$; thus, the average $\gamma$-factor is the same for the two directions, $\gamma_O(\bar{\beta}) = \gamma_O(\beta)$. However, the expressions do depend on the anisotropy that is interwoven within the formula of the average $\gamma$-factor, $\gamma_O(\beta;r)$. It is, then, useful to approximate the time-dilation and wavelength shifting for small anisotropies, $r \ll 1$,

$$\Delta t'_O / \Delta t = \lambda'_O / \lambda = \gamma_O(\beta;r) \cong \gamma_{r=0} + \gamma^3_{r=0} \beta^3 \cdot \tfrac{1}{2} r + O(r^2), \qquad (54)$$

thus, we observe a small increase of the time-dilation or wavelength shifting by $\delta \gamma_O / \gamma_O \cong \gamma^2_{r=0} \beta^3 \cdot \tfrac{1}{2} r$, with the isotropic $\gamma$-factor, $\gamma_{r=0} = (1-\beta^2)^{-1/2}$. This result may be measurable in high-precision experiments, which could set upper bounds on, or even determine, the level of anisotropy in the suggested relativistic theory. If the speed is significantly less than the light speed, $\beta \ll 1$, then, the isotropic $\gamma$-factor approximates to $\gamma_{r=0} - 1 \cong -\tfrac{1}{2}\beta^2$; the next approximation term is third order in terms of speed, $-\tfrac{1}{3}\beta^3 \cdot r$, as opposed to the fourth order for the isotropic case, $-\tfrac{1}{8}\beta^4$. Chou et al. [105] constructed a plot of measurements of $\Delta\lambda / \lambda$ against speed, with precision sufficient to show the second order dependence $\Delta\lambda / \lambda \sim -\tfrac{1}{2}\beta^2$, where $\Delta\lambda = \lambda'_O - \lambda$ denotes the wavelength shifting. Here we suggest that a high precision measurement of the next speed term could be decisive for the anisotropy value of $r$. As shown in Figure 6, the plot of $\Delta\lambda / \lambda + \tfrac{1}{2}\beta^2$ against speed will depend on the fourth or third power of speed, depending on the relativity if it is isotropic ($r=0$) or anisotropic ($r>0$), respectively.

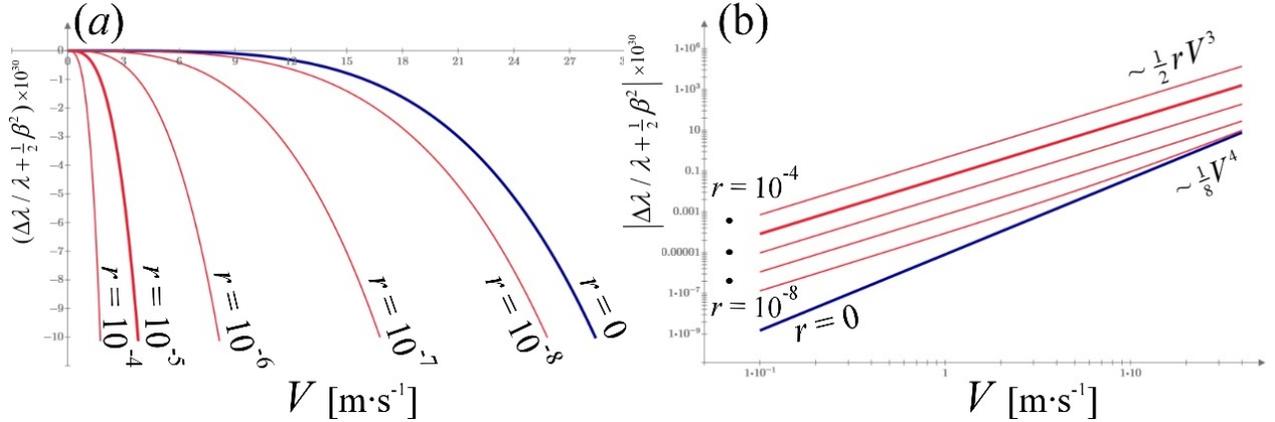

**Figure 6.** (a) Plot of the third speed term of the wavelength shifting, $\Delta\lambda / \lambda + \tfrac{1}{2}\beta^2$, as a function of speed $V$, for various values of anisotropy $r$. (b) Same plot on log-log scale, showing the different speed exponents.

## 8. Discussion and Conclusions

Thermodynamic relativity is a novel and broadened theoretical framework of special relativity, unified for describing entropies and velocities, and consistent with both thermodynamics and kinematics. The new theory should not be confused with some relativistic adaptation of thermodynamics; instead, it is a unification of the physical disciplines of thermodynamics and kinematics that share a identical description within the framework of relativity.

To achieve this framework, we first developed the most general entropic composability, which describes the possible ways the entropy of a system can be shared among its constituents. The partitioning of the total entropy into the entropies of the system's constituents is called *H*-partitioning, because entropies *S* are involved in the respective mathematical expression via a generic function, *H*(*S*). The partitioning function *H* has specific physical and mathematical properties for being consistent with thermodynamics, as identified in Section 2. The *H*-partitioning shows explicitly how two parts of entropy can be added to give the composed entropy. This is a generalization of the standard addition, which we call kappa-addition.

The *H*-partitioning and kappa-addition are interwoven with three physical characteristics of entropy: (1) There is no privileged reference frame with zero entropy (We note that setting zero entropy at zero temperature is only a convenient definition originally formulated in the classical BG statistical framework.); (2) existence of a maximum entropic value, invariant and fixed for all reference frames; and (3) existence of stationarity, namely, when original constituents that reside in stationary states merge to form the composed system, it also resides in a stationary state.

It is remarkable that the developed theory of relativity of entropies and the traditional relativity of velocities are concepts based on identical postulates. In particular, Einstein's special relativity, postulates that: (1) the laws of physics take the same form in all inertial frames of reference (principle of relativity); (2) the speed of light in free space has the same value in all inertial frames of reference (existence of an invariant velocity); and (3) existence of stationarity of velocity for all the observers. The first two postulates follow the classical paradigm, while the third was considered a trivial condition in Einstein's special relativity for velocities, but is clearly necessary for entropies.

Any comparison of observations between reference frames for either entropies or velocities requires a connection, through which information can be exchanged. We showed that the algebra of kappa-addition provides the formulation that characterizes this connection. Indeed, kappa-addition forms a mathematical group on the set of entropies or velocities, thus, the unified framework is supplied with the properties of symmetry and transitivity, necessary for the zeroth law of thermodynamics and characterization of stationarity.

Therefore, the unified framework of relativity postulates can be stated, as follows:

1. *First postulate: There is no absolute reference frame in which entropy or velocity are zero; entropy and velocity are relative quantities, connected with the properties of symmetry and transitivity, broadly defined under a general addition rule that forms a mathematical group on the sets of their allowable values.*

2. *Second postulate: There exists a nonzero, finite value of entropy and of velocity, which remains fixed (i.e., constant for all times) and invariant (i.e., constant for all observers), thus, it has the same value in all reference frames, and constitutes the upper limit of any entropy and velocity, respectively.*
3. *Third postulate: If a system is stationary in thermodynamics (i.e., constant entropy - zeroth law of thermodynamics) and/or kinematics (i.e., constant velocity – characterization of inertia) for a reference frame O, it will be stationary for all reference frames that are stationary for O.*

The systematic methodology of the theory of Thermodynamic Relativity describes the development of adaptations of relativistic formalism within the above postulates (Figure 7). Such a formalism starts with the selection of the partitioning function *H*. An example, which was examined throughout the paper in detail, is the anisotropic special relativity connected with the specific partitioning function noted with $H_a(x)$. The rationale of selecting this *H* function is that it constitutes the most general case that leads to linear Lorentz transformation, and it includes both the disciplines of nonextensive thermodynamics and Einstein's special relativity as extreme special cases. The next step was to formulate the kappa-addition and its algebraic properties. Finally, one may proceed via entropy and thermodynamics and/or via velocity and kinematics.

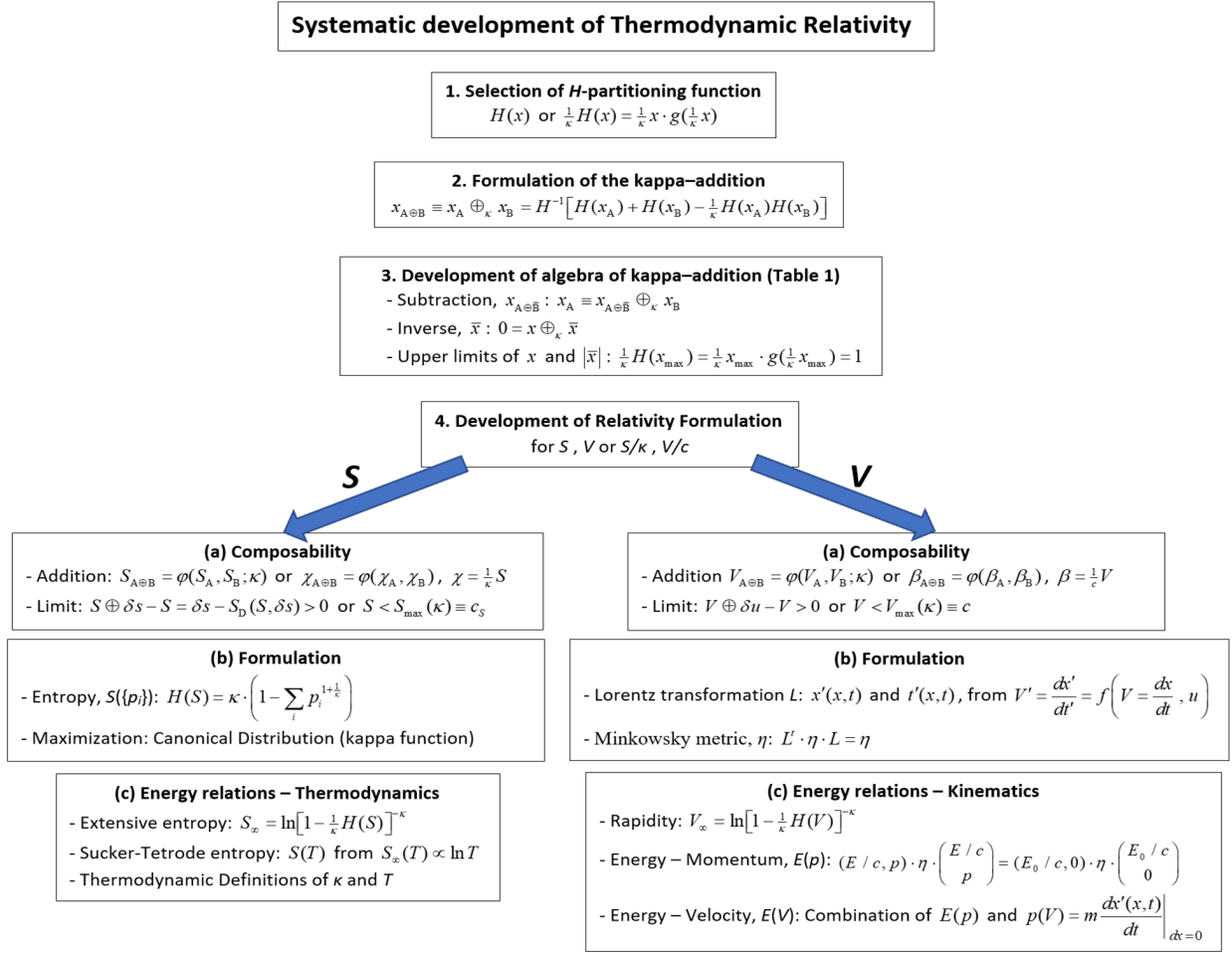

**Figure 7.** Systematic methodology of Thermodynamic Relativity theory.

Einstein's special relativity and nonextensive thermodynamics are two core foundations of fundamental physics. Thermodynamic relativity unifies the two disciplines in a theory that is consistent to both thermodynamics and kinematics. Moreover, space plasmas provide an observational ground truth in the development of a new and broader paradigm of theoretical physics. These are the natural laboratory for directly observing plasma particle distributions with long-rang interactions and thus correlations, and characterizing their thermodynamics; example is the thermodynamic characterization of the outer boundaries of our heliosphere via existing measurements from the Interstellar Boundary Explorer mission [106], and even more precise observations from the Interstellar Mapping and Acceleration Probe mission, launching in 2025 [107].

The paper developed the first attempt to generalize relativity in a thermodynamic context, leading naturally to the anisotropic and nonlinear relativity; (even for flat space, e.g., [108-111]). General relativity is a nonlinear theory due to the curved spacetime and leads to nonlinear transformations, while Einstein's special relativity is characterized by linear Lorentz transformations. Curved space is surely required with a nonlinear relativistic formalism. However, the theory of thermodynamic relativity developed in this study provides yet another natural way of constructing a nonlinear adaptation of relativity, that is, by adopting a nonlinear partitioning function $H$ such that the developed Lorentz transformation is nonlinear. Such an example can be one order higher partitioning function than $H_a(x) = x/(1+\frac{a}{\kappa}x)$ in Eq.(17), that is, $H_a(x) = x(1+\frac{b}{\kappa}x)/(1+\frac{a}{\kappa}x)$. Therefore, thermodynamic relativity constitutes a whole new path of generalization, as compared to the "traditional" passage from special to general theory based on curved spacetime geometries.

It is now straightforward to use the strength and capabilities of thermodynamics and kinematics through the unification of thermodynamic relativity, to (i) test the naturally derived version of anisotropic special relativity, (ii) apply both the thermodynamics and kinematics relativistic frameworks for describing particle populations in space and plasma science as well as entropy more broadly in many other disciplines, and (iii) extend the theory to further generalizations, such as, the nonlinear adaptation of relativity. In fact, this work obviously begs the questions: 1) Can the full capabilities of thermodynamic relativity, where the arbitrary $H$-partitioning corresponds to nonlinear transformations, describe and extend the general theory of anisotropic and nonlinear relativity, also in the presence of gravity? and 2) is the theory of thermodynamic relativity one step further in the ultimate unification of physics through particles' motions and connectedness with each other?

**Acknowledgements:** This work was funded in part by the IBEX mission as part of NASA's Explorer Program (80NSSC18K0237) and IMAP mission as a part of NASA's Solar Terrestrial Probes (STP) Program (80GSFC19C0027).

**Data availability:** All data generated or analyzed during this study are included in this published article.

# The theory of thermodynamic relativity

## (Supplementary)

George Livadiotis* and David J. McComas

Department of Astrophysical Sciences, Princeton University, Princeton, NJ 08544, USA

We introduce the theory of thermodynamic relativity, a unified theoretical framework for describing both entropies and velocities, and their respective physical disciplines of thermodynamics and kinematics, which share a surprisingly identical description with relativity. This is the first study to generalize relativity in a thermodynamic context, leading naturally to anisotropic and nonlinear adaptations of relativity; thermodynamic relativity constitutes a new path of generalization, as compared to the "traditional" passage from special to general theory based on curved spacetime. We show that entropy and velocity are characterized by three identical postulates, which provide the basis of a broader framework of relativity: (1) no privileged reference frame with zero value; (2) existence of an invariant and fixed value for all reference frames; and (3) existence of stationarity. The postulates lead to a unique way of addition for entropies and for velocities, called kappa-addition. We develop a systematic method of constructing a generalized framework of the theory of relativity, based on the kappa-addition formulation, that is fully consistent with both thermodynamics and kinematics. We call this novel and unified theoretical framework for simultaneously describing entropy and velocity "thermodynamic relativity". From the generality of the kappa-addition formulation, we focus on the cases corresponding to linear adaptations of special relativity. Then, we show how the developed thermodynamic relativity leads to the addition of entropies in nonextensive thermodynamics and the addition of velocities in Einstein's isotropic special relativity, as in two extreme cases, while intermediate cases correspond to a possible anisotropic adaptation of relativity. Using thermodynamic relativity for velocities, we start from the kappa-addition of velocities and construct the basic formulations of the linear anisotropic special relativity; e.g., the asymmetric Lorentz transformation, the nondiagonal metric, and the energy-momentum-velocity relationships. Then, we discuss the physical consequences of the possible anisotropy in known relativistic effects, such as, (i) matter-antimatter asymmetry, (ii) time dilation, and (iii) Doppler effect, and show how these might be used to detect and quantify a potential anisotropy.

**Keywords:** *Entropy; Relativity; Kappa distributions; Plasma*

## A. *Formulation of thermodynamic relativity for entropy*

The absolute and relative definitions of velocity and their distinction are clear enough physical concepts. The same holds with the formulation and measurements of the relative velocity. However, this is not the case with entropy. While we presented our interpretation of entropy as a relative physical quantity, the formulation of entropy is basically through an absolute definition. Here, we develop the formulation of relative entropy.

In its absolute definition, the entropy is given in terms of a single probability distribution $\{p_i\}$. Let the auxiliary argument $\Phi$, which is related to the extensive measure $S_\infty$, given in Eq.(5), as

$$\Phi \equiv 1 - \tfrac{1}{\kappa} H(S) = e^{-\tfrac{1}{\kappa} S_\infty} . \tag{A1}$$

Therefore, if we know Φ we can find entropy. We then derive Φ. As mentioned in Eq.(6), the *H*-partitioning can be written in the product form, $[1 - \frac{1}{\kappa} H(S_{A \oplus B})] = [1 - \frac{1}{\kappa} H(S_A)] \cdot [1 - \frac{1}{\kappa} H(S_B)]$. Then, the composition of the originally independent systems A and B leads to the simple product rule: $\Phi_{A \oplus B} = \Phi_A \cdot \Phi_B$. The corresponding probability distributions are $p^{A \oplus B} = p^A \cdot p^B$ and the only physically meaningful way the argument Φ can be expressed by these probabilities is through power-law relations (e.g., [57,58]), that is, e.g., $\Phi = \sum_i p_i^{1 + \frac{1}{\kappa}}$, where the summation is on the states, $\{p_i\} = p_1, p_2, \ldots$ Hence, given Eq.(A1), we find

$$1 - \tfrac{1}{\kappa} H(S) = \Phi = \sum_i p_i^{1 + \frac{1}{\kappa}} \quad \text{or} \quad H(S) = \kappa \cdot \left(1 - \sum_i p_i^{1 + \frac{1}{\kappa}}\right). \tag{A2}$$

Next, we continue with the formulation of relative entropy. Given the anisotropic case of *H*-partitioning function, $\Phi = 1 - \tfrac{1}{\kappa} H(S) = (1 + \tfrac{a-1}{\kappa} S)/(1 + \tfrac{a}{\kappa} S)$, and using the separate kappa for each direction, $\kappa_1$ and $\kappa_2$, we find $\Phi = (1 - \tfrac{1}{\kappa_1} S)/(1 + \tfrac{1}{\kappa_2} S)$. Then, we solve in terms of entropy,

$$S = \frac{1 - \Phi}{\frac{1}{\kappa_1} + \frac{1}{\kappa_2} \Phi} = \frac{1 - \Phi}{\frac{1}{\kappa} - \frac{1}{\kappa_2}(1 - \Phi)}. \tag{A3a}$$

with its inverse given by $\bar{S}$, for which $S \oplus_\kappa \bar{S} = 0$, namely

$$\bar{S} = -\frac{1 - \Phi}{\frac{1}{\kappa_2} + \frac{1}{\kappa_1} \Phi} = -\frac{1 - \Phi}{\frac{1}{\kappa} - \frac{1}{\kappa_1}(1 - \Phi)}. \tag{A3b}$$

We note that the inverse entropy has the same relation with $\Phi^{-1}$ as the entropy with $\Phi$,

$$\bar{S} = \frac{1 - \Phi^{-1}}{\frac{1}{\kappa_1} + \frac{1}{\kappa_2} \Phi^{-1}} = \frac{1 - \Phi^{-1}}{\frac{1}{\kappa} - \frac{1}{\kappa_1}(1 - \Phi^{-1})}, \tag{A4}$$

namely,

$$\bar{S}(\Phi) = S(\Phi^{-1}). \tag{A5}$$

For two systems A and B, the entropy of B as measured from A, $S_{B,A}$, and the entropy of A as measured from B, $S_{A,B}$, are inverse to each other, i.e., $S_{B,A} \oplus_\kappa S_{A,B} = 0$, where $S_{A,B} = \bar{S}_{B,A}$. The respective Φ arguments must have the following relationship: if $S_{B,A} = S(\Phi_{B,A})$, then $S_{A,B} = S(\Phi_{B,A}^{-1})$, i.e.,

$$\Phi_{A,B} = \Phi_{B,A}^{-1} \quad \text{for} \quad S_{B,A} \oplus_\kappa S_{A,B} = 0. \tag{A6}$$

Then, we adopt the division of two different Φ's, in order to capture the relativity correspondence between two systems, as shown in Eq.(A6). For this, we take again the two systems A and B, with respective probability distributions $\{p_i\}$ and $\{q_i\}$, and the positive and negative directions: $\{q_i\} \underset{\kappa_2}{\overset{\kappa_1}{\rightleftarrows}} \{p_i\}$. Then, the argument Φ of B measured by A, is given by

$$\Phi = \frac{\sum\limits_i p_i (p_i / q_i)^{\frac{1}{\kappa_1}}}{\sum\limits_i q_i (p_i / q_i)^{\frac{1}{\kappa_2}}}. \tag{A7}$$

This form is supplied with the desired properties of the relative entropy: (1) Inverse entropy relationship (validation of Eq.(A6)); (2) Recovery of the absolute formulation of kappa entropy; and (3) Entropy of the observer's own system is zero.

1) This expression of $\Phi$ generalizes the absolute definition in Eq.(A2) to capture the relative definition, namely, the relation $\Phi(\{q_i\},\kappa_2;\{p_i\},\kappa_1) = \Phi(\{p_i\},\kappa_1;\{q_i\},\kappa_2)^{-1}$ that is valid for relative entropies, as shown in Eq.(A6). Indeed, we have

$$\Phi_{B,A} = \frac{\sum_i p_i (p_i/q_i)^{\frac{1}{\kappa_1}}}{\sum_i q_i (q_i/p_i)^{\frac{1}{\kappa_2}}}, \quad \Phi_{A,B} = \frac{\sum_i q_i (q_i/p_i)^{\frac{1}{\kappa_2}}}{\sum_i p_i (p_i/q_i)^{\frac{1}{\kappa_1}}} = \Phi_{B,A}^{-1}, \tag{A8a}$$

and

$$S_{B,A} = \frac{1-\Phi_{B,A}}{\frac{1}{\kappa}-\frac{1}{\kappa_2}(1-\Phi_{B,A})}, \quad S_{A,B} = \overline{S}_{B,A} = \frac{1-\Phi_{B,A}^{-1}}{\frac{1}{\kappa}-\frac{1}{\kappa_2}(1-\Phi_{B,A}^{-1})} = \frac{1-\Phi_{A,B}}{\frac{1}{\kappa}-\frac{1}{\kappa_2}(1-\Phi_{A,B})}. \tag{A8b}$$

2) The absolute definition of entropy recovers the kappa entropy. This is derived from setting any of the directional kappa to infinity, i.e., (i) $\kappa_2 \to \infty, \kappa_1 = \kappa$, or (ii) $\kappa_1 \to \infty, \kappa_2 = \kappa$. Namely, when the upper limit becomes infinite at one direction, the entropy recovers the standard kappa formalism, while when both kappas tend to infinity, then, the relative entropy coincides with the Kullback–Leibler extensive definition [69]:

$\kappa_2 \to \infty, \kappa_1 = \kappa$:

$$S = \kappa \cdot \left[1 - \sum_i p_i (p_i/q_i)^{\frac{1}{\kappa}}\right] \xrightarrow{\kappa\to\infty} -\sum_i p_i \ln(p_i/q_i), \tag{A9a}$$

and

$\kappa_1 \to \infty, \kappa_2 = \kappa$:

$$\overline{S} = \kappa \cdot \left[1 - \sum_i q_i (q_i/p_i)^{\frac{1}{\kappa}}\right] \xrightarrow{\kappa\to\infty} -\sum_i q_i \ln(q_i/p_i). \tag{A9b}$$

3) Entropy of the observer's system is zero. Indeed, for a system (let this be noted with A) observing and describing itself, we obtain the expected result of zero entropy. Indeed, $\Phi(\{p\}=\{q\}) = \Phi^{-1} = 1$, thus, $S(\{p\}=\{q\}) = \overline{S}(\{p\}=\{q\}) = 0$; hence, $\Phi_{A,A} = 1$ and $S_{A,A} = 0$.

**B. *Formulation of thermodynamic relativity for velocity***
**B.1. Kappa Addition for anisotropic relativity**

Here we show that the addition rule of velocities/entropies that correspond to the anisotropic relativity.

The kappa addition of variable $x$, standing for entropy $x=S$ or for 1-dimensional velocity $x=V$, and for the partitioning function $H(x)$, is given by

$$H(x_{A+B}) = H(x_A) + H(x_B) - \tfrac{1}{\kappa}H(x_A) \cdot H(x_B), \tag{B1}$$

conveniently written in the product form

$$1 - \tfrac{1}{\kappa}H(x_{A+B}) = [1 - \tfrac{1}{\kappa}H(x_A)] \cdot [1 - \tfrac{1}{\kappa}H(x_B)]. \tag{B2}$$

Let the partitioning function

$$H(x) = \frac{x}{1+\frac{a}{\kappa}x}. \tag{B3}$$

We have $\frac{a}{\kappa}H(x) = \frac{a}{\kappa}x/(1+\frac{a}{\kappa}x)$, thus $1-\frac{1}{\kappa}H(x) = (1+\frac{a-1}{\kappa}x)/(1+\frac{a}{\kappa}x)$. Substituting in (B2), we derive

$$\frac{1+\frac{a-1}{\kappa}x_{A+B}}{1+\frac{a}{\kappa}x_{A+B}} = \frac{1+\frac{a-1}{\kappa}x_A}{1+\frac{a}{\kappa}x_A} \cdot \frac{1+\frac{a-1}{\kappa}x_B}{1+\frac{a}{\kappa}x_B}. \tag{B4}$$

Solving in terms of $x_{A+B}$,

$$x_{A+B} = \frac{1-X}{\frac{1-a}{\kappa} + \frac{a}{\kappa} \cdot X}, \text{ with } X \equiv \frac{1-\frac{1-a}{\kappa}x_A}{1+\frac{a}{\kappa}x_A} \cdot \frac{1-\frac{1-a}{\kappa}x_B}{1+\frac{a}{\kappa}x_B}, \text{ or} \tag{B5}$$

$$x_{A+B} = \frac{(1+\frac{a}{\kappa}x_A)(1+\frac{a}{\kappa}x_B) - (1-\frac{1-a}{\kappa}x_A)(1-\frac{1-a}{\kappa}x_B)}{\frac{1-a}{\kappa}(1+\frac{a}{\kappa}x_A)(1+\frac{a}{\kappa}x_B) + \frac{a}{\kappa}(1-\frac{1-a}{\kappa}x_A)(1-\frac{1-a}{\kappa}x_B)} = \frac{x_A + x_B - \frac{1-2a}{\kappa}x_A x_B}{1 + \frac{a(1-a)}{\kappa^2}x_A x_B} . \tag{B6}$$

Extreme cases:

- $a=0$, Simple entropy defect, for describing nonextensive statistical mechanics.

$$x_{A+B} = x_A + x_B - \frac{1}{\kappa}x_A x_B, \ x \to S, \tag{B7a}$$

- $a=1/2$, Einstein's isotropic relativity.

$$x_{A+B} = \frac{x_A + x_B}{1 + \frac{1}{(2\kappa)^2}x_A x_B}, \ x \to V, \ 2\kappa \to c, \tag{B7b}$$

while the case of $a=1$, leading to reversed entropy defect, $x_{A+B} = x_A + x_B + \frac{1}{\kappa}x_A x_B$, is unphysical (see Figure 3 of the main text).

**B.2. Asymmetric Lorentz Transformation Matrix**

We start with a body of moving at velocity $V$, as measured from a reference frame O. If this frame moves with a velocity $u$ as measured by a reference frame O´, then, the velocity $V'$, measured by O´, is given by the kappa addition of velocities:

$$V' = \frac{V(1 - r\frac{1}{c_0}u) + u}{1 + (1 - \frac{1}{4}r^2)\frac{1}{c_0^2}uV} . \tag{B8}$$

The above kappa addition of velocities generalizes the Galilean standard addition, $V' = V + u$.

Using this addition of velocities, we can find the corresponding Lorentz transformation. We express the velocities as $V = dx/dt$, $V' = dx'/dt'$, and set the linear transformation,

$$\begin{aligned} dx' &= A dx + B dt, \\ dt' &= C dx + D dt, \end{aligned} \tag{B9}$$

where we will find the involved constants in terms of $u$, as follows

$$V' = \frac{dx'}{dt'} = \frac{A dx + B dt}{C dx + D dt} = \frac{(A/D)V + (B/D)}{(C/D)V + 1} . \tag{B10}$$

Comparing Eqs.(B8,B10), we find

$$A/D = 1 - r\frac{1}{c_0}u, \ B/D = u, \ C/D = (1-\frac{1}{4}r^2)\frac{1}{c_0^2}u . \tag{B11}$$

The fourth equation that determines $D$, comes from the determinant of the a-Lorentz transformation, which is set to 1, i.e., $AD - BC = 1$, or

$$D^{-1} = \sqrt{(A/D) - (B/D)(C/D)} = \sqrt{(1-\frac{1}{2}r\beta)^2 - \beta^2} , \tag{B12}$$

where we recall the normalized velocity $\beta \equiv \frac{1}{c_0} u$. For zero anisotropy this recovers the gamma-factor of Einstein's special relativity, $\gamma = \sqrt{1-\beta^2}$. Then, we define $\gamma_r \equiv D^{-1}$, thus,

$$\gamma_r = 1/\sqrt{(1-\tfrac{1}{2}r\beta)^2 - \beta^2}, \text{ or } \gamma_r = 1/\sqrt{(1-u/c_1)\cdot(1+u/c_2)}. \tag{B13}$$

Therefore, we express the Lorentz transformation, after integrating Eq.(B10) and substituting Eqs.(B11,B13),

$$x' = \gamma_r \cdot \left[(1-r\tfrac{1}{c_0}u)x + u\cdot t\right],$$
$$t' = \gamma_r \cdot \left[(1-\tfrac{1}{4}r^2)\tfrac{1}{c_0^2}u\cdot x + t\right], \tag{B14}$$

or, in a matrix form

$$\begin{pmatrix} c_0 t' \\ x' \end{pmatrix} = L_r(\beta) \cdot \begin{pmatrix} c_0 t \\ x \end{pmatrix}, \; L_r(\beta) \equiv \gamma_r \cdot \begin{pmatrix} 1 & (1-\tfrac{1}{4}r^2)\beta \\ \beta & 1-r\beta \end{pmatrix}, \tag{B15}$$

where $L_r(\beta)$ denotes the anisotropic Lorentz transformation, and it is an asymmetric matrix.

The inverse transformation is given by the Lorentz transformation of the inverse velocity,

$$L_r^{-1}(\beta) = \gamma_r \cdot \begin{pmatrix} 1-r\beta & -(1-\tfrac{1}{4}r^2)\beta \\ -\beta & 1 \end{pmatrix} = \overline{\gamma}_r (1-r\beta)^{-1} \cdot \begin{pmatrix} 1-r\beta & (1-\tfrac{1}{4}r^2)\overline{\beta}(1-r\beta) \\ \overline{\beta}(1-r\beta) & (1-r\beta)(1-r\overline{\beta}) \end{pmatrix}$$
$$= \overline{\gamma}_r \cdot \begin{pmatrix} 1 & (1-\tfrac{1}{4}r^2)\overline{\beta} \\ \overline{\beta} & 1-r\overline{\beta} \end{pmatrix} = L_r(\overline{\beta}). \tag{B16}$$

where the inverse $\gamma$-factor is:

$$\overline{\gamma}_r \equiv \gamma_r(\overline{\beta}) = \gamma_r \cdot (1-r\beta). \tag{B17}$$

**B.3. Properties of the γ-factor**

We show the following:
- Expression in terms of directional light speeds $c_1$ and $c_2$:

After calculus with Eq.(B13), we obtain $\gamma_r = \sqrt{[1-(1+\tfrac{1}{2}r)\beta]\cdot[1+(1-\tfrac{1}{2}r)\beta]}$, and using the directional light speeds, shown in Eq.(30), we find

$$\gamma_r = \sqrt{(1-u/c_1)\cdot(1+u/c_2)}, \tag{B18}$$

where the formalism naturally distinguishes the speed of light at the two directions.
- Inverse gamma:

$$\overline{\gamma}_r \equiv \gamma_r(\overline{\beta}) = \frac{1}{\sqrt{(1-\overline{u}/c_1)\cdot(1+\overline{u}/c_2)}} = \frac{1-(\tfrac{1}{c_1}-\tfrac{1}{c_2})u}{\sqrt{(1-u/c_1)\cdot(1+u/c_2)}} = \gamma_r \cdot [1-(\tfrac{1}{c_1}-\tfrac{1}{c_2})u] = \gamma_r \cdot (1-r\beta). \tag{B19}$$

We can easily derive the identity:

$$\overline{\gamma}_r \cdot \overline{\beta} = -\gamma_r \cdot \beta, \tag{B20}$$

which is used in Section 6.6 to show the conservation of momentum.
- Directional gamma:

After some calculus with the negative velocity, shown in Eq.(34), we derive

$$1-\tfrac{1}{c_1}\overline{V} = \frac{1+\tfrac{1}{c_2}V}{1-(\tfrac{1}{c_1}-\tfrac{1}{c_2})V}, \; 1+\tfrac{1}{c_2}\overline{V} = \frac{1-\tfrac{1}{c_1}V}{1-(\tfrac{1}{c_1}-\tfrac{1}{c_2})V}, \tag{B21}$$

from where we extract the directional gamma-factors:

$$\gamma_1 \equiv 1/\sqrt{(1-\tfrac{1}{c_1}\overline{V})(1-\tfrac{1}{c_1}V)} = 1/\sqrt{(1+\tfrac{1}{c_2}\overline{V})(1+\tfrac{1}{c_2}V)} \equiv \gamma_2, \tag{B22}$$

and

$$\frac{1-\tfrac{1}{c_1}\overline{V}}{1+\tfrac{1}{c_2}\overline{V}} = \frac{1+\tfrac{1}{c_2}V}{1-\tfrac{1}{c_1}V}, \tag{B23}$$

which will be used in Section 7.3 to show the Doppler effect.

**B.4. Non-diagonal Metric**

Here we show the derivation of the Minkowsky metric, that is a non-diagonal matrix, generalizing the diagonal matrix *diag*(-1,1) that corresponds to the isotropic Einstein's special relativity. Take two events ($ct_1, x_1$) of ($ct_2, x_2$). Note that for the sake of simplicity, we consider two-dimensional spacetime, throughout this paper, one dimension for time and one spatial dimension, along which the motion is set to take place. The distance $\Delta s$ between those two events is given by $\sqrt{-(ct_2-ct_1)^2+(x_2-x_1)^2}$, instead of the standard magnitude of $\sqrt{(ct_2-ct_1)^2+(x_2-x_1)^2}$; the former determines the spacetime length via the diagonal Minkowsky metric of *diag*(-1,1), while the latter is the spacetime length via the diagonal Euclidean metric of *diag*(1,1), i.e.,

$$\begin{aligned}(ct_2-ct_1 \quad x_2-x_1)\begin{pmatrix}-1 & 0\\ 0 & 1\end{pmatrix}\begin{pmatrix}ct_2-ct_1\\ x_2-x_1\end{pmatrix} &= -(ct_2-ct_1)^2+(x_2-x_1)^2 = \Delta s^2,\\ (ct_2-ct_1 \quad x_2-x_1)\begin{pmatrix}1 & 0\\ 0 & 1\end{pmatrix}\begin{pmatrix}ct_2-ct_1\\ x_2-x_1\end{pmatrix} &= (ct_2-ct_1)^2+(x_2-x_1)^2 \neq \Delta s^2.\end{aligned} \tag{B24}$$

The Minkowsky metric is a matrix broadly denoted with $\eta$; here, we use the notions of $\eta_0$ for the isotropic and $\eta_r$ for the anisotropic cases. The metric models the spacetime in special relativity, and it is used to derive the spacetime distance between any two events. In the above, the metric was used in its simplest form, which is in the Cartesian coordinates, that is, the diagonal matrix, *diag*{-1,1}. Then, the invariant infinitesimal spacetime length $ds^2$, or proper time $c^2d\tau^2 = -ds^2$, is given by

$$-c^2d\tau^2 = ds^2 = (cdt \quad dx)\begin{pmatrix}-1 & 0\\ 0 & 1\end{pmatrix}\begin{pmatrix}cdt\\ dx\end{pmatrix} = -(cdt)^2+dx^2, \tag{B25}$$

which is better known in the closed form of $ds^2 = d\vec{x}^t \cdot \eta_0 \cdot d\vec{x}$, with the spacetime vector $d\vec{x}$ and its transpose vector $d\vec{x}^t$, defined by

$$d\vec{x} = \begin{pmatrix}cdt\\ dx\end{pmatrix},\ d\vec{x}^t = \begin{pmatrix}cdt\\ dx\end{pmatrix}^T = (cdt \quad dx). \tag{B26}$$

The Lorentz transformation is formally defined through its property to keep invariant the spacetime length $ds^2$ or proper time $c^2d\tau^2 = -ds^2$,

$$-c^2d\tau^2 = ds^2 = (cdt' \quad dx') \cdot \eta_0 \cdot \begin{pmatrix}cdt'\\ dx'\end{pmatrix} = (cdt \quad dx) \cdot \eta_0 \cdot \begin{pmatrix}cdt\\ dx\end{pmatrix}, \tag{B27}$$

where the Lorentz transformation matrix $L_0$, and its transpose $L_0^T$, are involved as follows:

$$\begin{pmatrix}cdt'\\ dx'\end{pmatrix} = L_0(\beta) \cdot \begin{pmatrix}cdt\\ dx\end{pmatrix},\ (c_0dt' \quad dx') = (cdt \quad dx) \cdot L_0^T(\beta). \tag{B28}$$

Hence, substituting Eq.(B28) in Eq.(B27), we find

$$(cdt \quad dx)L_0^T(\beta)\cdot\eta_0\cdot L_0(\beta)\cdot\begin{pmatrix}cdt\\dx\end{pmatrix}=(cdt \quad dx)\cdot\eta_0\cdot\begin{pmatrix}cdt\\dx\end{pmatrix}, \tag{B29}$$

which holds for any vector, if

$$L_0^T(\beta)\cdot\eta_0\cdot L_0(\beta)=\eta_0. \tag{B30}$$

Then, we can use Eq.(B30) to derive the respective metric $\eta_0$, as a matrix with determinant equal to -1. The isotropic Lorentz transformation has its simplest form in the 1-dimensional case, i.e.,

$$L_0(\beta)=\gamma_0(\beta)\cdot\begin{pmatrix}1 & \beta\\ \beta & 1\end{pmatrix}, \quad \gamma_0(\beta)\equiv 1/\sqrt{1-\beta^2}. \tag{B31}$$

It can be easily verified that Eq.(B31) obeys Eq.(B30) when $\eta_0 = diag\{-1,1\}$; substituting (B31) in (B30), we derive the metric to be $\eta_0 = diag\{-1,1\}$.

Moreover, in the anisotropic adaptation of relativity, we consider the observable average light speed $c_0$ to describe the spacetime vectors

$$d\vec{x}=\begin{pmatrix}c_0 dt\\dx\end{pmatrix}, \quad d\vec{x}^t=\begin{pmatrix}c_0 dt\\dx\end{pmatrix}^t=(c_0 dt \quad dx). \tag{B32}$$

The Lorentz transformation is again defined through the invariance of the spacetime length $ds$ or proper time $d\tau$, with $c_0^2 d\tau^2 = -ds^2$,

$$-c_0^2 d\tau^2 = ds^2 = (c_0 dt' \quad dx')\cdot\eta_r\cdot\begin{pmatrix}c_0 dt'\\dx'\end{pmatrix}=(c_0 dt \quad dx)\cdot\eta_r\cdot\begin{pmatrix}c_0 dt\\dx\end{pmatrix}, \tag{B33}$$

where

$$\begin{pmatrix}c_0 dt'\\dx'\end{pmatrix}=L_r(\beta)\cdot\begin{pmatrix}c_0 dt\\dx\end{pmatrix}, \quad (c_0 dt' \quad dx')=(c_0 dt \quad dx)\cdot L_r^t(\beta). \tag{B34}$$

Hence, we now find

$$(c_0 dt \quad dx)L_r^t(\beta)\cdot\eta_r\cdot L_r(\beta)\cdot\begin{pmatrix}c_0 dt\\dx\end{pmatrix}=(c_0 dt \quad dx)\cdot\eta_r\cdot\begin{pmatrix}c_0 dt\\dx\end{pmatrix}, \tag{B35}$$

or

$$L_r^t(\beta)\cdot\eta_r\cdot L_r(\beta)=\eta_r. \tag{B36}$$

Using the anisotropic Lorentz transformation matrix, as shown in Eq.(B15), we derive the form of anisotropic metric, $\eta_r$. We find:

$$\begin{pmatrix}1 & \beta\\ (1-\tfrac{1}{4}r^2)\beta & 1-r\beta\end{pmatrix}\begin{pmatrix}A & \sqrt{AC+1}\\ \sqrt{AC+1} & C\end{pmatrix}\begin{pmatrix}1 & (1-\tfrac{1}{4}r^2)\beta\\ \beta & 1-r\beta\end{pmatrix}=\gamma_r^{-2}\cdot\begin{pmatrix}A & \sqrt{AC+1}\\ \sqrt{AC+1} & C\end{pmatrix}, \tag{B37}$$

where we set the nondiagonal term accordingly, so that $\det(\eta_r)=-1$. Then, we derive the four dependent equations:

$$\begin{aligned}&(1-\gamma_r^{-2})\cdot A+\beta^2\cdot C+2\beta\cdot\sqrt{AC+1}=0,\\ &(1-\tfrac{1}{4}r^2)\beta\cdot A+(1-r\beta)\beta\cdot C+[(1-r\beta)+(1-\tfrac{1}{4}r^2)\beta^2-\gamma_r^{-2}]\cdot\sqrt{AC+1}=0,\\ &(1-\tfrac{1}{4}r^2)^2\beta^2\cdot A+[(1-r\beta)^2-\gamma_r^{-2}]\cdot C+2(1-r\beta)(1-\tfrac{1}{4}r^2)\beta\cdot\sqrt{AC+1}=0,\end{aligned} \tag{B38}$$

and substituting $\gamma_r$ from (B13), we find the following three equations:

$$[r + (1 - \tfrac{1}{4}r^2)\beta] \cdot A + \beta \cdot C + 2 \cdot \sqrt{AC + 1} = 0,$$
$$(1 - \tfrac{1}{4}r^2) \cdot A + (1 - r\beta) \cdot C + 2(1 - \tfrac{1}{4}r^2)\beta \cdot \sqrt{AC + 1} = 0, \qquad (B39)$$
$$(1 - \tfrac{1}{4}r^2)^2 \beta \cdot A + [-r + (1 + \tfrac{3}{4}r^2)\beta] \cdot C + 2(1 - r\beta)(1 - \tfrac{1}{4}r^2) \cdot \sqrt{AC + 1} = 0,$$

where the combination of the first equation with any of the other two equations to cancel the square root leads to only one independent equation, that is, $C = -A \cdot (1 - \tfrac{1}{4}r^2)$. Then, substituting this to any of the three equations (let's say the first), we find $\tfrac{1}{2}r \cdot (-A) = \sqrt{1 - A^2 \cdot (1 - (\tfrac{1}{2}r)^2)}$. Note that the coefficient $A$ must be negative so that to be consistent with the isotropic case ($r=0$) where $A = -1$; hence, the square root has only a positive sign. Interestingly, the solution of this last equation is that $A$ is independent of $r$, i.e., $A(r) = -1$. Finally, we end up with

$$\eta_r = \begin{pmatrix} -1 & \tfrac{1}{2}r \\ \tfrac{1}{2}r & 1 - \tfrac{1}{4}r^2 \end{pmatrix}. \qquad (B40)$$